\documentstyle[12pt]{article}
\begin{document}
\pagestyle{plain}
\newcommand{\be}{\begin{equation}}
\newcommand{\ee}{\end{equation}}
\newcommand{\bea}{\begin{eqnarray}}
\newcommand{\eea}{\end{eqnarray}}
\newcommand{\vp}{\varphi}
\newcommand{\pr}{\prime}
\newcommand{\sech} {{\rm sech}}
\newcommand{\cosech} {{\rm cosech}}
\newcommand{\psib} {\bar{\psi}}
\newcommand{\cosec} {{\rm cosec}}
\def\vs {\vskip .3 true cm}
\centerline {\bf Fractional Statistics and Chern-Simons Field}
\centerline {\bf Theory in 2+1 Dimensions}
\vs
\centerline {\bf Avinash Khare$^*$}
\centerline {Institute of Physics, Sachivalaya Marg,}
\centerline {Bhubaneswar 751005, India.}
\centerline {Email: khare@iopb.res.in}
\vs
{\bf Abstract}

The question of anyons and fractional statistics in field theories in 2+1 
dimensions with 
Chern-Simons (CS) term 
is discussed in some detail. Arguments are spelled out 
as to why fractional statistics is only possible in two space dimensions. 
This phenomenon is most naturally discussed within the framework of field
theories with CS term, hence as a prelude to this discussion 
I first discuss the
various properties of the CS term. In particular 
its role as a gauge field mass term is emphasized. In the presence of the
CS term, anyons can appear in two different ways i.e. either as 
soliton of the corresponding 
field theory or as a fundamental quanta carrying 
fractional statistics and both approaches are elaborated in some detail.

\vfill
* To be Published in INSA (Indian National Science Academy) Book 2000
\eject

\section{\bf Introduction} 

Many of us have wondered some time or the other if one can have 
nontrivial science 
and technology in two 
space dimensions; but the usual feeling is that two space dimensions
do not 
offer enough scope for it. This question, to the best of my knowledge, was
first addressed in 1884 
by E.A. Abbot in his satirical novel {\it Flatland} 
\cite{1ab}. The first serious
book on this topic appeared in 1907 entitled {\it An episode of  
Flatland} \cite{1hi}. In this
book C.H. Hinton 
offered glimpses of the possible science and technology in the
flatland. A nice 
summary of these two books appeared as a chapter entitled 
{\it Flatland} in a book in 1969 edited by 
Martin Gardner \cite{1ga}. Inspired by this summary, in 1979 A.K. Dewdney 
\cite{1de} published a 
book which contains several laws of physics, chemistry, 
astronomy and biology in the flatland. However, all 
these people 
missed one important case where physical laws are much more complex,
 nontrivial and hence 
interesting in the flatland than in our three dimensional
world. I am 
referring here to the case of quantum statistics. In last two decades 
it has been realized that whereas in three and higher space dimensions all
particles must 
either be bosons or fermions (i.e. they must have spin of $n\hbar$
or $(2n+1)\hbar /2$ with n=0,1,2,... 
and must obey Bose-Einstein or Fermi-Dirac
statistics respectively), 
in two space dimensions the particles can have any fractional
spin and can 
satisfy $\it any$ fractional statistics which is interpolating between the
two. The particles 
obeying such 
statistics are generically called as $\it anyons$ \cite{ak}.  
In other words, if one takes one 
anyon slowly around the other then in general the phase 
acquired is $exp(\pm i\theta$).
 If $\theta$ =0 or $\pi$ (modulo $2\pi$) 
then the particles are bosons or fermions
respectively 
while if $0 < \theta < \pi$ then the particles are termed as anyons. 

From our experience with fermions and bosons it is well known that the 
question of spin and statistics can be properly handled only within the 
formalism of relativistic quantum field theory. Thus it is of interest to
enquire if one can also understand the ideas of anyons and fractional statistics within the formalism of relativistic quantum field theory. This is the issue
that we would like to discuss in this article.

Before I go into the details, one might wonder 
if our discussion is merely of academic interest? 
The answer to the question is no. 
In fact it is a 
surprising fact that two, one and even zero dimensional experimental
physics is possible in our three-dimensional world.
This is because of the third law of thermodynamics,
which states that all the 
degrees of freedom freeze out in the limit of zero temperature,
it is possible to strictly 
confine the electrons to surfaces, or even to lines or points.
Thus it may happen that 
in a strongly confining potential, or at sufficiently low 
temperatures, 
the excitation 
energy in one or more directions may be much higher than the average 
thermal 
energy of the 
particles, so that those dimensions are effectively frozen out.
Of course, 
even then, at the basic level, the fundamental particles 
are certainly fermions or bosons. However, the most direct and appropriate
discussion 
of the low energy behavior of a material is usually in terms of the 
quasi-particles. 
The hope is that at least in some of these cases the quasi-particles
could be anyons. This hope 
has in fact been realized in the case of the fractionally
quantized Hall effect 
where the quasi-particles are believed to be charged vortices 
i.e. charged
anyons \cite{1la}. Recent experiments \cite{1cl} seem to confirm
the existence of fractionally 
charged excitations and hence indirectly of anyons.  

The plan of the article is the following. In Sec.II, I first spell out as
to why fractional statistics is only possible in two space dimensions.
It turns out that 
the phenomenon 
of fractional statistics is most naturally discussed within the 
framework of field theories with CS term.
As a prelude to this discussion, in Sec.III,  
I discuss the various properties of the CS term. 
In particular its role as a gauge field mass term and its 
behavior under the discrete 
transformations of parity (P) and time-reversal (T) is emphasized. 
In the presence of the CS term, 
anyons can appear in two different 
ways (i.e. either as soliton of the corresponding field theory or as 
fundamental quanta carrying fractional statistics) 
and both 
approaches are elaborated in some detail in the next three sections. 
The charged vortex   
solutions in Abelian Higgs model with CS term 
are obtained in Sec.IV, and it is pointed out that these charged vortices 
represent the first relativistic model for (extended) charged anyons. 
I also
construct the 
charged vortex solutions in pure CS theory in both the
relativistic and the non-relativistic  settings. 
In Sec.V, I discuss an example
of neutral relativistic anyons by considering the soliton solutions in the
$CP^1$ model with the Hopf term which is one of 
the $\it avtars$ of the CS term.
Finally, in Sec.VI, I 
elaborate upon the other approach in which fundamental 
fields of theories with CS term themselves carry fractional 
spin and obey fractional statistics. 

\section{\bf Why Anyons in Only Two Dimensions?}

Before we come to the question of fractional statistics, it might be 
worthwhile to understand as to why unlike in three and higher space 
dimensions, the eigenvalue of the spin angular momentum operator can 
take any fractional
value in units of $\hbar$.
The point is that the spin in two dimensions differs fundamentally from
the spin in higher dimensions. This is because whereas in three and
higher space dimensions, the spin angular momentum algebra is
non-commutative i.e.  
\be\label{2.1}
[ S_i, S_j ] = i\hbar \varepsilon_{ijk} S_k \, ; \ \ i,j,k = 1,2,3
\ee
in two space dimensions, it is a trivial commutative algebra since
only one generator (say $S_3$) is available which obviously commutes
with itself. As a result, there is no analogue of the quantization of
the angular momentum, which arises in three and higher space dimensions
from the nonlinear commutation relation (\ref{2.1}). Here 
$\varepsilon_{ijk}$ is the completely antisymmetric tensor.

Now, in relativistic quantum field theory, there is a deep and profound
connection between the spin and the statistics i.e. particles with half
integer spin are fermions, satisfying Fermi-Dirac statistics, while those 
with integer spin are bosons, satisfying Bose-Einstein statistics.
This immediately suggests that in two dimensions the particles
may exhibit fractional (i.e. any) statistics. 
In a remarkable paper Leinaas and Myrheim \cite{2lm} showed that this
is indeed so.
Before we come to a proper discussion about the statistics, it is worth
clarifying as to what exactly one means by quantum statistics. In most
text books on statistical mechanics, the term ``quantum statistics''
refers to the phase picked up by a wave function when two identical
particles are interchanged, i.e, under the permutation of the
particles. But this is slightly misleading and has been correctly
criticized in the literature \cite{2mi}. If
the particles are {\it strictly identical}, the word permutation has
no physical meaning since a given configuration and the one obtained
by the permutation of the 
particle coordinates are merely two different ways
of describing the {\it same} particle configuration. 
The term quantum statistics actually
refers to the phase that arises when two particles are
adiabatically transported giving rise to the exchange. In this book,
we shall be concentrating on this definition of quantum statistics.
It is a coincidence that in three and higher dimensions, the two
definitions, based on the permutation and the adiabatic exchange of two
particles, coincide, but in two dimensions the two definitions give very
different answers. 

The key reason for the fractional statistics in two
dimensions is the principle of indistinguishability of
identical particles. It is one of the most important characteristics
of quantum mechanics (vis a vis classical mechanics) and it has
profound physical consequences. The principle is in fact older than
quantum mechanics. It was introduced by John Willard Gibbs even in
classical statistical mechanics to resolve the famous Gibbs paradox.
Even though this principle has been with us for a very long time, 
unfortunately, its full significance was not appreciated till 1977
and that is how one missed the possibility of fractional statistics
in two dimensions for all these years.  

Following Leinaas and Myrheim \cite{2lm}, let us enquire about
 the configuration space of a system of identical particles ? Normally
one considers the full phase space in statistical mechanics but it
turns out that configuration space is enough for this discussion.
Suppose one particle space is $X$. Then what is the configuration space
of $N$ identical particles ? The Naive answer is $X^N$, which, even
though true locally, is {\it not correct} globally. Why? The reason
is, since the particles are strictly identical, hence there is no
distinction between the points in $X^N$ that differ only in the ordering
of the particle coordinates. For example, consider the point
\be\label {2.9}
{\bf x} = ({\bf x}_1, {\bf x}_2,..., {\bf x}_N)
\ee 
in $X^N$ where ${\bf x}_i\in X$ for $i = 1,2,...,N.$ Now consider
another point ${\bf x}'$ in $X^N$ which is obtained from $\bf x$ by the 
permutation $p$ of the particle indices i.e.
\be\label{2.10}
{\bf x}' = P({\bf x}) = 
({\bf x}_{P^{-1}(1)},..., {\bf x}_{P^{-1}(N)}) \, .
\ee
Clearly, both describe the {\it same} physical configuration of the
system. Thus the true configuration of the $N$-particle system is {\it
not} $X^N$ but it is the space $X^N/S_N$ which is obtained by
identifying points in $X^N$ that represent the same physical
configuration, i.e. it is obtained from $X^N$ by dividing out by the
action of the symmetry group $S_N$. Note that $S_N$ is a discrete,
finite group obtained by permutation of N identical particles. As a
result, the space $X^N/S_N$ is locally isomorphic to $X^N$ except at
its singular points. However, the global properties of the two spaces
are very different. Whereas $X^N$ has only regular points when $X$ is
regular, those points in $X^N/S_N$ which correspond to a coincidence
of the positions of two or more particles are in fact singular points
of $X^N/S_N$. Thus to calculate the configuration space of identical
particles, such singular points must be excluded by say hard-core
constraint so that we can determine if two particles have been
exchanged or not. This of course does not make much difference
classically. However, in the quantum case the global properties of
the configuration space are of deep significance and this results in
the possibility of fractional statistics. It is worth emphasizing 
that this is the crux of
the whole 
matter and it is this fact which was missed for about fifty years!

It turns out that the removal of such singular point in two space 
dimensions makes the space multiply connected while for three and
higher space dimensions it is still doubly connected. 
That is why, in two dimensions it is possible to define paths that wind
around the origin an arbitrary number of times counted with orientation.
As a consequence, when one quantizes a system of identical particles then
one can show that in two dimensions it is possible to consistently assign 
{\it any value} to the phase arising due to the exchange of two identical
particles.
Since in two dimensions one can distinguish the clockwise winding from the
anti-clockwise winding, hence without any loss of generality one can
assign the phases $e^{+i\theta}$ and $e^{-i\theta}$ respectively, in the 
case of 
the anti-clockwise and the clockwise windings.

At this point, it 
may be worthwhile to mention few key properties of anyons.
\begin{enumerate}
\item Anyons must necessarily 
violate the discrete symmetries of parity (P)
and time reversal (T) if $0 < \theta < \pi$ since the clockwise and the 
anti-clockwise windings have different phase factors.
\item Anyons are sort of in between the bosons and the fermions i.e. the
repulsion between two anyons in the ground state monotonically increases 
as $\theta$ goes from 0 to $\pi$
with there being no repulsion between two bosons. Thus, 
in a sense, anyons are closer to the fermions than to the bosons since 
all of them will satisfy a generalized form of Pauli
exclusion principle. 
\item It turns out that whereas the 
permutation group which is at the heart of the Bose-Einstein and the
Fermi-Dirac statistics, it is the braid group which is at the heart of the
fractional statistics. In 
particular, whereas there are two one dimensional
representations of the 
permutation group (the identical one and the alternating
one, corresponding to the Bose-Einstein and Fermi-Dirac statistics 
respectively), the braid group admits a continuous parameter family of one
dimensional 
representations which one usually identifies with the parameter
$\theta$ which characterizes fractional statistics.
\item Is there a relation between the anyonic statistics and the 
parastatistics ? The answer is {\it no}. They are built on  
two different structures i.e. whereas the  
Parastatistics corresponds to
the higher dimensional representation of the permutation group while
anyons correspond to the one dimensional representation of
the braid group. 
\end{enumerate}

{\bf Quantum Statistics in One Dimension} 

Since we have been talking about 
the possible quantum statistics in various 
dimensions, hence it may 
be worthwhile to also talk about the various possibilities
in one dimension. Recall that the notion of the spin 
does not exist in one dimension since
there is no axis to rotate about in that case. Similarly the
concept of the quantum 
statistics is not uniquely defined in one dimension since the
position of two particles cannot be interchanged without their
passing through one another. As a result, the intrinsic statistics is
inextricably mixed up with the local interactions. In fact this
ambiguity is at the heart of the bosonization technique which allows
the same particle to be represented alternatively by a boson or a fermion
field. If, however, statistics is defined in terms of the exclusion
principle rather than the exchange of identical particles, then it is
possible to define quantum statistics in even one dimension \cite{2ha}. 

\section{Introduction to Chern-Simons Term}

We now want to 
understand how anyons occur in field theory. It turns out that
this is possible provided 
the CS term or its incarnation, the
Hopf term 
are present. It may therefore be worthwhile to first introduce the 
CS term (in 2+1 dimensions) and discuss its various 
properties \cite{6ak}. 
 
{\bf What is Chern-Simons Term?}

Consider the Lagrangian density for classical electrodynamics in 3+1 
dimensions as given by 
\be\label{6.1.1}
{\cal L} = - {1\over 4} F_{\mu\nu} F^{\mu\nu} 
+\overline\psi (i\gamma_{\mu}D^{\mu}-m)\psi
\ee
where $F_{\mu\nu} =\partial_{\mu}A_{\nu}-\partial_{\nu}A_{\mu}$ and 
$D_{\mu}=\partial_{\mu} - ie A_{\mu}$ is the covariant derivative. 
This Lagrangian is invariant under the local gauge transformation
\be\label{6.1.2}
\psi (x)\rightarrow e^{ie\alpha(x)} \psi(x)\, , 
\ A_{\mu} (x) \rightarrow A_{\mu} (x) +\partial_{\mu} \alpha (x) \, .
\ee
Similarly, for massless fermions (m=0), this Lagrangian is also invariant 
under the (global) chiral transformation
\be\label{6.1.3}
\psi (x)\rightarrow e^{i\gamma_5\beta}\psi (x)\, , 
\ A_{\mu} (x)\rightarrow A_{\mu} (x) \, .
\ee
The naive expectation was that, these two symmetries i.e. the 
gauge and the chiral
symmetries, which are valid at the classical level, will continue to hold 
good even in the 
quantum theory. As a consequence, one expected that the vector and the 
axial
vector currents
$j_{\mu} = \overline\psi \gamma_{\mu} \psi$ and 
$j^5_{\mu} = \overline\psi \gamma_{\mu}\gamma_5\psi$
which are 
conserved at the classical level, will continue to remain conserved
even in the quantum theory.
It has however, been shown that this is not so.
There is {\it no} 
regularization which can simultaneously preserve both these 
symmetries
at the quantum level. 
Because of the unexpected result, it was called an anomaly 
at that time (and unfortunately even today it is called so), even though 
the correct name should have been quantum mechanical symmetry breaking. 
Remarkably, the entire effect comes only from one loop diagram and 
two and higher loops do not
contribute to the 
anomaly. In view of our strong faith in the gauge symmetry, 
one therefore says that it is the
chiral symmetry which is 
broken by the one loop quantum corrections. 
In particular, there is a gauge 
singlet
(axial) anomaly 
in any even dimension, ($2n$) so that the divergence of the 
gauge singlet axial current, 
even for
massless fermions, is non-zero and proportional to the corresponding 
Chern-Pontryagin (CP) density $P_{2n}$ in that (even) dimension $2n$ i.e. 
\be\label{6.1.4}
\partial^{\mu} j^5_{\mu} (x) \ \propto \ P_{2n} \, .
\ee
It is also well known that the 
CP Density can always be written as a total
divergence
\be\label{6.1.5}
P_{2n} = \ \partial_{\mu} \Lambda^{\mu} \, , \ \mu = 0,1,2,..., 2n-1 \, .
\ee
The object $\Lambda^{\mu}$, for a particular value of $\mu$ 
(say $\mu $ = 2n-1) naturally lives in odd $(2n-1)$
dimensions and is known as the CS density in that 
dimension. Thus, whereas 
the CP 
density lives in even space-time dimensions, the CS density 
lives in odd space-time 
dimensions. For
example, the gauge singlet anomaly in 3+1 dimensional quantum 
electrodynamics is given by 
\be\label{6.1.6}
\partial^{\mu} j_{\mu}^5 
= {e^2\over 2\pi} \varepsilon_{\mu\nu\lambda\sigma} F^{\mu\nu} 
F^{\lambda\sigma} 
= {e^2\over\pi} \partial^{\mu} (\varepsilon_{\mu\nu\lambda\sigma} A^{\nu} 
F^{\lambda\sigma})
\ee
so that the Abelian CS term in 2+1 dimensions is given by
\be\label{6.1.7}
J_{CS} = \int {\cal L}_{CS} \, d^3 x \ \propto \ \int d^3 x \, 
\varepsilon_{\nu\lambda\sigma} A^{\nu} 
F^{\lambda\sigma} \, .
\ee
Throughout this book we shall mainly be concerned with this CS 
term or its non-Abelian
generalization. Let us therefore discuss in some detail the 
various properties 
of this term.

{\bf Gauge Invariant Mass Term}

Let us consider pure electrodynamics in the presence of the Chern-Simons 
term in 2+1 dimensions \cite{6sho,6djt}
\be\label{6.2.1}
{\cal L} = - {1\over 4} F_{\mu\nu} F^{\mu\nu} 
+ {\mu\over 4} \varepsilon^{\mu\nu\lambda} F_{\mu\nu} A_{\lambda} \, .
\ee
Since the mass dimension of $A_{\mu}$ is $1/2$, hence it follows that the 
parameter $\mu$
has the dimension of mass. 
The field equation following from this Lagrangian can be written as 
\be\label{6.2.3}
(g^{\mu\nu} +{1\over\mu} \varepsilon^{\mu\nu\alpha} \partial_{\alpha} ) {^*F_{\nu}} = 0
\ee
where $^*F_{\nu}$ is the dual field strength which is a vector in 2+1 
dimensions
i.e.
\be\label{6.2.4}
^*F_{\nu}= {1\over 2} \varepsilon_{\nu\alpha\beta} F^{\alpha\beta} \, ; 
\ F_{\mu\nu} 
= \varepsilon_{\mu\nu\alpha} {^*F^{\alpha}} \, .
\ee
We thus find that, unlike 
the CP 
term which has only a nontrivial topology but no dynamics 
(being a total divergence), the CS term has nontrivial 
topology as well as
dynamics in it. 
On operating by $(g_{\beta\eta}
-{1\over \mu} \varepsilon_{\beta\eta\delta} \partial^{\delta})$
to Eq. (\ref{6.2.3}), we get
\be\label{6.2.5a}
(\Box + \mu^2) {^*F_{\beta}} = 0
\ee
which clearly shows that the gauge field excitations are massive with 
the gauge field
mass $\mu$ being the coefficient of the CS term. 
We have thus shown that the CS term when added to the Maxwell  
term, acts 
as the {\it gauge invariant gauge field mass term}. It is worth 
adding that
this remarkable property of having a 
gauge invariant mass  
term for the gauge
field in the action itself is very special to 2+1 dimensions. 

{\bf Behavior Under C, P, and T}

Let us consider the behaviour of the CS term as well as the 
Dirac Lagrangian
\be\label{6.3.1}
{\cal L}_D = i\overline\psi (\gamma_{\mu}\partial^{\mu}-m)\psi
\ee
under the discrete transformations $C$ (charge conjugation), $P$(parity) and $T$ 
(time reversal).
Here, $\psi$ is a two 
component
spinor with mass $m ( > 0)$ and the mass dimension of $\psi$ is $1$. We use the 
following two-dimensional realization of the Dirac algebra
\be\label{6.3.2}
\gamma^0 =\sigma^3 \, , \ \gamma^1 
= i\sigma^1 \, , \ \gamma^2 = i\sigma^2 \, ,
\ee
\be\label{6.3.3}
\gamma^{\mu}\gamma^{\nu} = g^{\mu\nu} 
- i\varepsilon^{\mu\nu\alpha} \gamma_{\alpha} \, ; 
\ g^{\mu\nu} = diag. (1,-1,-1)
\ee
where $\sigma^i$ are the usual Pauli matrices.

It is easily shown that under charge conjugation 
\be\label{6.3.4}
C A_{\mu} C^{-1} = - A_{\mu} \, , \ C\psi C^{-1} = \sigma^1 \psi^+
\ee
so that the action is invariant under $C$. On the other hand, under parity
transformation, the gauge and the Fermi fields transform as follows
\be\label{6.3.5}
P A^{0,2} (t,{\bf r}) P^{-1} = A^{0,2} (t,{\bf r}') \, , 
\ P A^1 (t,{\bf r}) P^{-1} 
= - A^1 (t,{\bf r}') \, , 
\ee
\be\label{6.3.6}
P \psi (t,{\bf r}) P^{-1} = \sigma^1 \psi (t,{\bf r}') \, .
\ee
Note that in 2+1 dimensions, 
the parity transformation is somewhat unusual i.e.
${\bf r} = (x,y), \ {\bf r}' = (-x,y)$ (or $(x,-y)$). On the other hand, $(-x,-y)$ 
corresponds to 
rotation (and not space
reflection). As a result, we find that the mass terms 
for both
the Fermi and the gauge fields (i.e. $m\overline\psi\psi$ and the 
CS term) are not invariant under parity. 
Similarly, time-inversion changes the signs of both the mass terms since
\be\label{6.3.7}
T A^0 (t,{\bf r}) T^{-1} = A^0 (-t,{\bf r}) \, , \ T {\bf A} (t,{\bf r}) T^{-1} 
= - {\bf A} (-t,{\bf r}) \, , 
\ee
\be\label{6.3.7a}
T \psi (t,{\bf r}) T^{-1} = \sigma^2 \psi (-t,{\bf r}) \, . 
\ee
Thus, both the CS  term as well as the fermion mass term, 
$m\overline\psi\psi$ are non-invariant under $P$ as well
as $T$. However, they are invariant under the combined operation $PT$ and hence 
the $CPT$ 
symmetry is still valid. Note that in $3+1$ dimensions though, 
$m\overline \psi\psi$ is invariant 
under $P, C$ and $T$ separately.

Finally, let us talk about the photon spin. One can show that the 
CS photon spin is
$1(-1)$ if CS mass $\mu > 0 (< 0)$ while the spin of the massless 
photon is zero. Further,
in either case, the photon has only one degree of freedom. 

{\bf Coleman - Hill Theorem}

It turns out that because of the $P$ and $T$ 
violating but gauge invariant CS term,  
the most general form for the vacuum polarization tensor consistent with 
Lorentz
and gauge invariance is more general than in other dimensions i.e.
\be\label{6.4.1}
\Pi_{\mu\nu} (k) = (k^2 g_{\mu\nu}-k_{\mu} k_{\nu}) \Pi_1 (k^2) 
- i \varepsilon_{\mu\nu\lambda} k^{\lambda}\Pi_2 (k^2) \, .
\ee
Note that the second term 
on the right hand side is odd under $P$ and $T$. It is 
clear that any
$P$ and $T$ violating interaction will contribute to $\Pi_2(k^2)$. For example, the
fermion mass term which violates both $P$ and $T$, does contribute to 
$\Pi_2(k^2)$ at one
loop. Remarkably enough, it was discovered that at two loops, however, there is 
no
contribution to $\Pi_2 (0)$ and hence to Chern-Simons mass \cite{6ak}. 
Inspired by 
this result, Coleman and Hill \cite{6ch} have in fact
proved under very general conditions that $\Pi_2(0)$ receives no 
contribution from
two and higher loops in any gauge and Lorentz invariant theory including 
particles of
spin 1 or less (An open question is whether this is also valid for higher spin
theories, specially spin-$3/2$). They only require that the 
matter fields be massive so 
that one does
not have to worry 
about the infrared problems. Further, they also assume that 
no part of the
free electro-magnetic Lagrangian density is hiding in the matter part of 
the Lagrangian.
It may be noted that their result is valid even for non-renormalizable 
interactions in
the presence of the gauge and Lorentz invariant regularization. 

Coleman and 
Hill
also claimed that at one loop, the only contribution to $\Pi_2(0)$ can 
come from the fermion loop.
This is, however, incorrect. In particular, there is no reason why 
$P$ and $T$ violating interactions
involving spin-0 or spin-1 particles should not contribute to 
$\Pi_2(0)$ at one loop.
In fact, it has been shown that the parity violating spin-0 \cite{6kmp} 
as well as
spin-1 interactions \cite{6hpr} do contribute to $\Pi_2 (0)$ at one loop.

{\bf Magneto-Electric Effect}

There are many crystals in nature like chromium oxide, which show the 
magneto-electric effect i.e.,
they also get magnetically polarized in an electric field and electrically
polarized in a magnetic field \cite{6ll,6od}. It is
well known that this effect depends upon having a $CP$-asymmetric medium. 
Mathematically, 
the signal for the magneto-electric effect in 2+1 dimensions
is that the relation between the excitation fields ${\bf D}$ and $H$ and 
${\bf E}$ and $B$
is modified to 
\be\label{6.5.1}
D_i = \chi_{ij}^{(e)} E_j + \chi_i^{(em)} B \, ; 
\ H = \chi^{(m)} B + \chi^{(me)}_i E_i \, .
\ee
It has been shown \cite{6kp} that the vacuum of the $2+1$
dimensional quantum electrodynamics with CS term 
also shows the magneto-electric 
effect. In particular, 
it has been shown that both $\chi_i^{(em)}$ and $\chi^{(me)}_i$ are
non-zero and 
proportional to $k_i\Pi_2 (k^2)$. Of course this is not really surprising
if one remembers that the CS term violates the discrete symmetries 
$P$ and $T$.

{\bf Chern-Simons Term by Spontaneous Symmetry Breaking}

We have seen above that the CS term 
provides mass to the gauge field. Now, usually
the gauge 
field mass is generated by spontaneous symmetry breaking; hence it is
worth enquiring whether the CS term can also be generated by 
spontaneous symmetry breaking. 
The answer to the question
is yes \cite{6pk}. This is because, unlike other
dimensions, in the $2+1$ case, 
one can have a more general definition of the 
covariant derivative. 
In particular, it is easily seen that
\be\label{6.6.1}
{\cal D}_{\mu} \psi = (\partial_{\mu} - ie A_{\mu} - ig \varepsilon_{\mu\nu\lambda} F^{\nu\lambda})\psi
\ee
also transforms as a covariant derivative, since the field
strength $F^{\nu\lambda}$ 
by itself is gauge invariant. Obviously, the same thing is also
true for a spin-$0$ 
charged scalar field. Now consider the following generalized
Abelian Higgs model in $2+1$ dimensions 
\be\label {6.6.2}
{\cal L} = -{1\over 4} F_{\mu\nu} F^{\mu\nu} 
+ {1\over 2} ({\cal D}_{\mu}\phi)^* ({\cal D}^{\mu}\phi)
-\alpha
(\mid\phi\mid^2-a^2)^2
\ee
where the generalized covariant 
derivative is as given by Eq. (\ref{6.6.1}). On 
expanding the 
term ${1\over 2} ({\cal D}_{\mu}\phi)^*  ({\cal D}^{\mu}\phi)$, 
we have 
\bea\label{6.6.3}
{ 1 \over 2} ({\cal D}_{\mu}\phi)^* ({\cal D}^{\mu}\phi) 
 = {1\over 2} (\partial_{\mu}
+ieA_{\mu})\phi^*(\partial^{\mu}
-ie A^{\mu})\phi
+{g^2\over 4} F_{\mu\nu}
F^{\mu\nu}\mid\phi\mid^2  \nonumber\\ 
 + ig \, {^*F_{\mu}}(\phi^*\partial^{\mu}\phi 
- \phi\partial^{\mu}\phi^*)  
 + eg\varepsilon_{\mu\nu\lambda}(\partial^{\mu}A^{\nu}) A^{\lambda} 
\mid\phi\mid^2
\eea
so that if $\phi$ acquires a nonzero vacuum expectation value then the 
Abelian
CS term is generated from the last term of this equation. Clearly a 
similar mechanism
should also work for the 
non-Abelian case, but technically it is a tougher problem
since one also 
has to generate the non-linear term. 

{\bf Lorentz Invariance From Gauge Invariance}

One of the remarkable properties of the Abelian CS term is 
that in this case
the Lorentz invariance of the action automatically follows from the gauge
invariance. In contrast, notice that the most general form of the
gauge invariant Maxwell Lagrangian in classical electrodynamics in $3+1$ 
dimensions is 
\be\label{6.7.1}
{\cal L} = {\bf E}^2 + a {\bf B}^2 \, .
\ee
It is only the demand of the Lorentz invariance which 
tell us that $a = - 1$ 
(In the $2+1$ case, 
$B$  is a pseudo scalar but the same argument is still valid). On the
other hand, if one writes the CS action as
\be\label{6.7.2}
I_{CS} = \int d^3 x    
[ \varepsilon_{ij} E^i A^j + a B A^0 ] \, , 
\ee
then the demand of the invariance of $I_{CS}$ 
under the gauge transformation
$A_{\mu}\rightarrow A_{\mu}+\partial_{\mu}\alpha$ 
fixes $a$ and uniquely gives us the 
CS action
which is automatically also Lorentz invariant.

{\bf Quantization of Chern-Simons Mass}

Let us now discuss the CS term in the non-Abelian gauge theories. We 
shall mention only
those properties which are special to the non-Abelian CS term. To 
begin with, 
notice that
the non-Abelian CS term
has an extra term compared to the Abelian 
case i.e.
\be\label{6.8.1}
I^{(CS)}_{na} = {\mu\over 4} \int d^3x \, \varepsilon^{\mu\nu\lambda} 
\, tr (F_{\mu\nu} A_{\lambda}
-{2\over 3} A_{\mu}A_{\nu}A_{\lambda})
\ee
where $A_{\mu}$ and $F_{\mu\nu}$ are matrices
\be\label{6.8.2}
A_{\mu} = g T^a A^a_{\mu} \, ; \ F_{\mu\nu} = g T^a F^a_{\mu\nu} = \partial_{\mu}
A_{\nu}-\partial_{\nu}A_{\mu}+[A_{\mu},A_{\nu}] \, .
\ee
Here, $T^a$ are the representation matrices of the gauge group $G$ satisfying 
\be\label{6.8.3}
[T^a, T^b ] = f^{abc} T^c
\ee
where $f^{abc}$ are the structure constants of the group. In the case of 
$SU(2)$, 
$T^a = \tau^a/2i$.

Let us 
now consider a non-Abelian gauge theory with the Chern-Simons term as given by  
\be\label{6.8.4}
{\cal L}_{na} = {1\over 2g^2} tr (F^{\mu\nu} F_{\mu\nu}) - {\mu\over 2g^2} 
\varepsilon^{\mu\nu\lambda}
tr (F_{\mu\nu}A_{\lambda}-{2\over 3}A_{\mu}A_{\nu}A_{\lambda})
\ee
As in the Abelian case, it is easily shown that the 
CS term provides a gauge invariant gauge field mass $\mu$. 

As in the Abelian case, the non-Abelian CS Lagrangian density 
changes by a total
derivative under an infinitesimal local gauge transformation so that the
corresponding action is invariant under such a gauge transformation.
However, the 
CS action is not invariant under finite (also called homotopically
non-trivial, or those which are not continuously deformable to the 
identity) gauge transformations as given by
\be\label{6.8.9}
A_{\mu}\longrightarrow U^{-1} A_{\mu} U + U^{-1} \partial_{\mu} U \, .
\ee
As a result, one finds that the action corresponding to the Lagrangian 
(\ref{6.8.4}) 
transforms as follows
\bea\label{6.8.10}
I_{na} & \longrightarrow & I_{na} 
 + \mu \int d^3 x \, \varepsilon^{\mu\nu\lambda} 
\, tr \bigg (\partial_{\nu} 
[A_{\mu}(\partial_{\lambda}U)U^{-1}] \bigg ) \nonumber\\
       & + & {\mu\over 3} \int d^3 x \, \varepsilon^{\mu\nu\lambda} \, 
tr \bigg [(\partial_{\mu} U)U^{-1} 
(\partial_{\nu} U)U^{-1}(\partial_{\lambda}U)U^{-1} \bigg ] \, .
\eea

Let us consider those gauge transformations which tend to the identity at 
temporal and 
spatial infinity 
so as to avoid a convergence problem i.e.
\be\label{6.8.11}
U(X) \stackrel{x\rightarrow\infty}{\longrightarrow} \ I \, .
\ee
It is now easily seen that the gauge field dependent surface integral 
in Eq. (\ref{6.8.10})
vanishes. However, the last term in the integral is non-zero. It can be 
converted to a 
surface integral once the integrand is rewritten as a total derivative. 
This can be
made manifest by using an explicit parameterization for $U$. For example, 
in the case of
$SU(2)$ (more generally, we choose $SU(2)$ 
sub-group of the gauge group $G$; for 
reasons that will
be clear soon), one can make use of the exponential parameterization
$U(X) = \exp (i\sigma^a\theta^a (x))$.
In this 
way one can show that under large gauge transformations, $I_{na}$   
is not invariant
but transforms as 
\be\label{6.8.13}
I_{na} \rightarrow I_{na} + {8\pi^2 \mu\over g^2} \omega (U)
\ee
where
\be\label{6.8.14}
\omega (U) = {1\over 24\pi^2}\int d^3 x \, \varepsilon^{\mu\nu\lambda} 
\, tr \bigg [ (\partial_{\mu} U)U^{-1} (\partial_{\nu}
U)U^{-1}(\partial_{\lambda}U)U^{-1} \bigg ]
\ee
is the winding number of the gauge transformation $U$. In particular,
if the gauge group $G$ is such that the third homotopy group of $G$ is 
non-trivial i.e.
\be\label{6.8.15}
\pi_3 (G) = Z
\ee
where $Z$ is the 
additive group of integers, then under these so called large 
gauge transformations, the action transforms as
\be\label{6.8.16}
I_{na}\rightarrow I_{na} +{8\pi^2\mu\over g^2} m
\ee
where $m$ is 
an integer. Note in particular, that Eq. (\ref{6.8.15}) is true for 
any 
gauge group $G$ of 
which $SU(2)$ is a sub-group. 
However, in the
path integral 
formulation, the action itself may or may not be gauge invariant 
but, 
it is the exponential of the action ($\exp (iI_{na}$)) which should be 
gauge
invariant. In this way we conclude that the non-Abelian gauge theory with 
the CS 
term
does not make sense in $2+1$ dimensions unless the CS mass 
$\mu$ is quantized \cite{6djt} 
in units
of $g^2/4\pi$ i.e. ($n=0, \pm 1, \pm 2,...$)
\be\label{6.8.17}
{8\pi^2 \mu\over g^2} = 2\pi n \ \ or \ \ \mu = {g^2\over 4\pi} n \, . 
\ee

This mass 
quantization is 
reminiscent of the famous Dirac quantization in the case of
magnetic monopole.
An important question to address is whether the quantization condition 
(\ref{6.8.17}) is 
respected by the 
quantum corrections. This issue was considered by Pisarski and 
Rao \cite{6pr}
for the case of a pure gauge theory (i.e. without any matter field). They 
found that
the quantization is indeed preserved to one loop; however, the integer on 
the right
hand side of Eq. (\ref{6.8.17}) is shifted by $N$ in case the gauge group 
$G = SU(N)$. 
Subsequently, it 
has been shown that there are no further corrections from 
two and higher loops in the
limit of the pure CS gauge theory \cite{6gi}.

How does the quantization condition modify in the presence of the 
matter fields? 
It has been shown that so long as the scalar field does not break the 
non-Abelian gauge 
symmetry, then the quantization condition remains unaltered. The 
massive fermions, 
of course, modify the quantization condition \cite{6pr} ;
the right hand side of Eq. ({\ref{6.8.17}) being shifted
by ${m_f\over \mid m_f\mid} T_R$, where $T_R$ is the 
Casimir generator for the 
gauge
group $G$ (i.e. $tr (T^a T^b) = -\delta^{ab} T_R)$, 
in case the fermions are 
in the 
fundamental representation of the gauge group $G$.
Thus the quantization is preserved so long as $T_R$ is an integer.

Much more 
interesting is the case of partial (spontaneous symmetry) breaking 
of a non-Abelian gauge symmetry. In this case it has been shown 
that if the non-Abelian
gauge symmetry $SU(N)$ is spontaneously broken to say $SU(M)\otimes U(1)$ 
(or even
several $U(1)'s)$, 
then the one-loop radiative correction to the right hand
side of the quantization condition (\ref{6.8.17}) \cite{6kmpp}
arises purely from the unbroken non-Abelian sector in question, the 
orthogonal $U(1)$
sector makes no contribution. This implies that the coefficient of the 
CS term is a
discontinuous function over the phase diagram of the theory.

{\bf Parity Anomaly}

Is our entire discussion about the CS term merely of academic 
interest ? Put 
differently,
some one might argue that since the CS term violates both the parity 
 and the time reversal invariance symmetries, 
why should one, in 
the first place, add such a term to the action ? The answer to this 
question, at least in
the non-Abelian gauge theories, is that even if one does not add the 
CS term to the
action at the 
tree level, it is automatically generated by the one loop radiative
corrections due to the so called parity anomaly \cite{6re}.
In particular, consider the action
\be\label{6.9.1}
I[A_{\mu},\psi] = \int d^3 x \, 
\bigg [ {1\over 2g^2} tr (F_{\mu\nu} F^{\mu\nu}) 
+i\overline\psi \gamma_{\mu} (\partial^{\mu}-ieA^{\mu})\psi \bigg ]
\ee
for an odd number of massless doublet of fermions 
in the fundamental representation 
coupled to $SU(2)$ gauge fields (more generally any gauge group $G$ of 
which $SU(2)$ is a sub-group
so that Eq. (\ref{6.8.15}) is 
satisfied; and the fermions are required only 
to be in the fundamental representation). 

This action 
is invariant under the gauge transformations (both large and small) 
as well as the  
discrete transformations of 
parity (P) and time reversal invariance (T). However, the 
effective action $I_{eff} [A]$, obtained
by integrating out the fermionic degrees of freedom, violates 
one of the two
symmetries. In other words, 
there is {\it no} regularization which can simultaneously
maintain the 
invariance of $I_{eff}[A]$ under the large gauge transformations
as well as $P$ and $T$. In view of the tremendous success of the gauge 
principle, one usually 
maintains the gauge invariance at the cost of the parity
and the time reversal invariance by
simply adding the CS term to the action (alternately one can also 
regulate it by using 
the $P$ and $T$ violating 
Pauli-Villars regularization). In this way, one finds that the 
CS term is induced by the 
radiative corrections even if it is absent at the tree level.
This is very similar to the way the CP term is induced in even
dimensions due to the gauge singlet (chiral) anomaly. 

{\bf Topological Field Theory}

One of the most remarkable property of the CS action is that it 
depends only on the 
antisymmetric tensor $\varepsilon_{\mu\nu\lambda}$ and not on the metric
tensor $g_{\mu\nu}$. As a result, the CS action in the flat and 
the curved
space is the same. Hence, the CS action, in both the Abelian and 
the non-Abelian cases, is an
example of the topological field theory \cite{6sc}. It might be 
mentioned here, that the
topological field theories give a natural framework for understanding the
Jones polynomials  of the Knot theory in terms of three dimensional terms.
Further, these theories have shed new light on conformal field theories in
two space-time dimensions.

Finally, the gravitational Chern-Simons term has also been considered 
\cite{6djt}
and shown to have some remarkable properties. In particular, whereas the 
massless
Einstein theory 
in $2+1$ dimensions is trivial, it acquires a propagating, 
massive, spin-$2$
degree of freedom when the CS term is present. Further, even 
though this 
topological term has third time derivative dependence, yet the theory is 
ghost-free
and unitary and one has a consistent quantum theory. 
The contribution of the topological mass term to the field equations also 
has a natural 
geometric significance: it is 
the three dimensional analogue of the Weyl tensor.

\section{Charged Vortex as Anyon in Field Theories}

In the last section, we have discussed in detail the various properties of
the CS term. In this section, we demonstrate the most dramatic effect
of this term i.e. the existence of charged vortex solutions
thereby
providing us with a relativistic model for the charged (extended) anyons. 

Before we discuss the charged vortex solutions, it might be worthwhile to 
mention how such solutions were historically discovered. A long time ago, 
Abrikosov \cite{7ab}
wrote down the 
electrically neutral vortex solutions in the Ginzburg-Landau 
theory which
is a mean-field theory of superconductivity. Subsequently, these vortices 
were experimentally
observed in the type-II superconductors. Nielsen and Olesen \cite{7no}
rediscovered these 
solutions in the context of the Abelian Higgs model which 
is essentially
a relativistic 
generalization of the Ginzburg-Landau theory. These people were
looking for string-like objects in relativistic 
field theory. It turns out that these 
vortices
have finite energy 
per unit length in $3+1$ dimensions (i.e. finite energy in $2+1$ 
dimensions as the 
vortex dynamics is essentially confined to the $x$-$y$ plane), 
quantized
flux, but are 
electrically neutral and have zero angular momentum. Subsequently, 
Julia
and Zee \cite{7jz} showed that the $SO(3)$ Gerogi-
Glashow model 
which admits t'Hooft-Polyakov monopole solution, also admits its
charged generalization i.e. 
the dyon solution with finite energy and finite, 
non-zero, 
electric charge. It was then natural for them to enquire whether the 
Abelian Higgs model, which admits
neutral vortex 
solutions with finite energy (in 2+1 dimensions), also admits 
its 
charged generalization or not. In the 
appendix of the same paper, Julia and Zee 
discussed
this question and 
showed that the answer is {\it no} i.e. unlike the monopole 
case, the Abelian
Higgs model does 
not admit charged vortices with finite energy and finite and
non-zero electric charge. 
More than ten years later, Samir Paul and I \cite{7pk}
showed that the Julia-Zee negative result can be overcome if one adds 
the CS term
to the 
Abelian Higgs model. In particular, we showed that the Abelian Higgs 
model
with CS term in 2+1 dimensions admits charged vortex solutions 
of finite energy and quantized, finite, 
Noether charge as well as flux. As an extra bonus, 
it was found
that these 
vortices also have non-zero, finite angular momentum which is in general
fractional. This 
strongly suggested that these charged vortices could in fact be
charged anyons which was subsequently rigorously shown by Fr\"ohlich and
Marchetti \cite{7fm}.

Strictly speaking, what one 
has obtained are the charged soliton solutions and not
the vortex solutions, 
but because of the close connection with the neutral vortex
solutions, one has continued to call them as charged vortices rather than
charged solitons. 

Consider an Abelian Higgs model with CS term as given by  
\be\label{7.1.1}
{\cal L} = -{1\over 4} F_{\mu\nu} F^{\mu\nu}
+{1\over 2} (D_{\mu}\phi)^* (D^{\mu}\phi)
- C_4 (\mid\phi\mid^2 - { C_2\over 2C_4})^2
+ {\mu\over 4} \varepsilon_{\mu\nu\lambda}F^{\mu\nu}A^{\lambda}
\ee
where $\mu$ is the Chern-Simons mass, $\phi$ denotes complex scalar field and 
$D_{\mu}\phi$ 
is the covariant derivative i.e.
\be\label{7.1.2}
D_{\mu}\phi = (\partial_{\mu} - ie A_{\mu})\phi \, .
\ee
Here $\phi, A_{\mu}$ as well as the 
gauge coupling constant $e$ have mass dimension
of $1/2$ while $C_4$ and $C_2$ have 
mass dimensions of $1$ and $2$ respectively.  
In order to 
obtain the charged vortex solutions, let us consider the following 
{\it ansatz}
\be\label{7.1.3}
{\bf A}({\bf x},t) = - e_{\theta} C_0 {(g(r)-n)\over r}, \ 
\phi ({\bf x},t) = C_0 f(r) e^{in\theta}, \ 
A_0({\bf x},t) = C_0 h(r)
\ee
where $g(r), h(r), f(r)$ 
are the dimension-less fields, r is the dimension-less 
length, while $C_0$ has mass dimension of $1/2$ i.e.
\be\label{7.1.4}
r = e C_0 \rho, \ C_0 = \sqrt{C_2/2 C_4} \, .
\ee
Note that $\rho$ and $\theta$ are related to $x$ and $y$ by $\rho = 
{\sqrt {x^2+y^2}}$
and $tan\theta = y/x$. It turns out 
that even though the Lagrangian (\ref{7.1.1}) 
has so many parameters, the dynamics essentially depends 
on two dimension-less variables, $\delta$ and $\lambda$ defined by
\be\label{7.1.5}
\lambda = {\sqrt {8 C_4/e^2}}, \ \delta = \mu/e C_0 \, .
\ee
The field equations which follow from here are
\be\label{7.1.6}
g''(r) - {1\over r} g' (r) - g f^2 = \delta r h'(r)
\ee
\be\label{7.1.7}
h'' (r) + {1\over r} h' (r) - hf^2 = {\delta\over r} g' (r)
\ee
\be\label{7.1.8}
f''(r) +{1\over r} f'(r) - {g^2f\over r^2} +{\lambda^2\over 2} f(1-f^2) = - fh^2
\ee
where $g'(r)\equiv {d g(r)/dr}$. The 
corresponding field energy can be shown to be 
\be\label{7.1.9}
E_n = \pi C^2_0 \int^{\infty}_0 rdr \bigg [ {1\over r^2} ({dg\over dr})^2
+({df\over dr})^2+
({dh\over dr})^2+h^2f^2+{g^2f^2\over r^2}+{\lambda^2\over 4}(1-f^2)^2\bigg ]
\ee

Several remarks are in order at this stage.
\begin{enumerate}
\item As expected, in the limit h = 0 (i.e. $A_0=0$) and $\delta = 0$ (i.e.
$\mu = 0$) the field equations reduce to those of the neutral vortex case.
From the Gauss 
law  Eq. (\ref{7.1.7}) it also follows that if $\delta$ (i.e. $\mu$) 
is non-zero, 
then $A_0$ {\it must} also be non-zero thereby justifying the {\it ansatz} 
(\ref{7.1.3}).
\item  The boundary conditions for finite energy solutions are 
\be\label{7.1.10}
\lim_{r\rightarrow \infty}  \ f(r) = 1, h(r) = 0 = g(r)
\ee
\be\label{7.1.11}
\lim_{r\rightarrow 0}  \ f(r) = 0, g(r) = n, h(r) = \beta
\ee
where $\beta$ is an arbitrary number while $n = 0, \pm 1, \pm 2$...\, .
\item  From these boundary conditions it immediately follows that the magnetic
flux is quantized in units of $2\pi/e$ i.e.
\be\label{7.1.12}
\Phi \equiv \int B d^2 x = - {2\pi\over e}\int^{\infty}_0 r d r 
({1\over r}{dg\over dr}) = {2\pi n\over e} \, .
\ee
It may be noted that even for the neutral vortices, the flux is quantized in 
units
of $2\pi\over e$. The underlying reason for the flux quantization 
is same in both the
cases i.e. both are topological objects with the underlying boundary 
conditions being such 
that there is a non-trivial mapping from the space time to the group 
manifold i.e.
$\pi_1 (U(1))=Z$, with $Z$ being the set of integers, 
forming a group under addition.
\item  From the 
Gauss law Eq. (\ref{7.1.7}), it then follows that these vortices 
also have a
non-zero and finite Noether charge which is quantized in units of 
$2\pi\mu/e$. This is easily seen by noting that
in terms of the electric and the magnetic fields, the Gauss law equation 
can be written as
\be\label{7.1.13}
{\bf \nabla} \cdot {\bf E} + \mu B = \rho
\ee
where $\rho$ is the Noether charge density. On integrating both sides of this
equation, it then follows that 
\be\label{7.1.14}
Q\equiv \int \rho d^2 x = \mu \int B d^2 x = {2\pi\mu\over e} n \, .
\ee
Note that $\int {\bf \nabla} \cdot {\bf E} d^2 x = 0$, since, 
because of the Higgs
mechanism, both ${\bf E}$ and B fall off exponentially at long distances.  
This is probably for the first time that the quantization of the
Noether charge has followed from  purely topological considerations. In a 
sense, 
relation (\ref{7.1.14}) can be 
looked upon as the (2+1)-analogue of the Witten
effect \cite{7wi}. Let us recall the work of Witten who had shown 
that in
the presence of the $CP$ and $T$ violating CP term, the 
t'Hooft-Polyakov
monopole acquires electric charge whose fractional 
part is proportional to the
coefficient of the CP term. It must however be remembered that 
whereas
the Witten effect is purely a quantum mechanical effect, in our case, 
the vortices acquire a non-zero charge at the classical 
level itself due to the presence of the CS term.
\item It is also clear from here that in the Abelian Higgs model 
(without the CS term),  one cannot
have vortices having 
simultaneously the finite energy as well as the finite, 
non-zero Noether 
charge. The point is, in the absence of the CS term, the Gauss law 
Eq. (\ref{7.1.13}) gives on integration
\be\label{7.1.15}
Q\equiv \int \rho \, d^2 x = \int {\bf \nabla} \cdot {\bf E} \, d^2 x \, .
\ee
The only way Q can be non-zero and finite is if 
there is a non-zero contribution
to the integral around $r\rightarrow 0$ i.e. if 
${\bf E} \rightarrow {1/r}$ as
$r\rightarrow 0$. But in that case, the electrical field energy 
$\int {\bf E}^2 d^2 x$ diverges logarithmically \cite{7jz}.
\item  The energy-momentum 
tensor $T_{\mu\nu}$ for this model can be obtained 
by varying the curved space form of the action with respect to the metric
\be\label{7.1.16}
T_{\mu\nu} = {1\over 2} (D_{\mu}\phi)^* (D_{\nu}\phi)
+{1\over 2} (D_{\nu}\phi)^* D_{\mu}\phi
-g_{\mu\nu} ({\cal L} 
- {\mu\over 4}\varepsilon_{\alpha\beta\gamma} F^{\alpha\beta} A^{\gamma})
\ee
where the Lagrangian $\cal L$ is as given by Eq. (\ref{7.1.1}). Note that the 
CS term, being 
independent
of the metric tensor $g_{\mu\nu}$, does not contribute to the energy 
momentum tensor $T_{\mu\nu}$. Using 
this $T_{\mu\nu}$
and the 
field equations, the angular momentum carried by the charged vortices 
can be shown to be
\be\label{7.1.17}
J \equiv \int d^2 x \, \varepsilon^{ij} \, x_i T_{oj} = - {nQ\over 2e} = - {\pi\mu\over e^2} n^2 
= - {Q\Phi\over 4\pi} \, .
\ee
Thus, unlike the neutral vortices, the 
angular momentum of the charged vortices
is non-zero and is solely determined by their 
charge  and 
flux. Besides, the angular 
momentum
of $n$ 
superimposed 
charged vortices is $n^2$ {\it and not $n$} times the angular 
momentum of a single vortex.
Further, since the 
CS mass $\mu$ is not quantized in the Abelian case, hence 
this angular momentum $J$ can in general take any fractional
value. This strongly suggests that these charged vortices are  
charged anyons. 
Fr\"ohlich and Marchetti \cite{7fm} 
have in fact rigorously proved that these  
charged vortices are charged anyons. 
In particular, they constructed quantum one vortex operator and then
evaluated the phase acquired when one such vortex is slowly taken 
round the other.
They also show that the charged vortices 
cannot be localized in bounded regions but are localized in space-like 
cones in three-dimensional Minkowski space-time \cite{7bf}.
Unfortunately their treatment is rather involved
and is beyond the scope of this pedagogical article. 
Thus the solitons of 
the Abelian Higgs model with the CS term  
provides us with a relativistic field theory model for the extended 
charged anyons. 
\item The magnetic moment of these 
vortices can be computed by using the field
equations 
and one can show that, whereas for the neutral vortices it is equal 
to the
flux $\Phi ( = 2\pi n/e)$, the charged ones acquire an extra contribution 
\be\label{7.1.18}
K_z \equiv \int ({\bf r} \times {\ J})_z \, d^2 x = {2\pi n\over e} 
+{2\pi\delta\over e} \int^{\infty}_0 rh (r) dr \, .
\ee
\end{enumerate}

{\bf Unusual Higgs Mechanism}

One must 
now solve the field Eqs. (\ref{7.1.6}) to (\ref{7.1.8}) and show the 
existence of the charged 
vortex solutions. 
To date, no analytic solution has been obtained of these 
field
equations. However, it is 
easily seen that for large $r$, the asymptotic values
of the gauge and the Higgs fields are reached exponentially fast
\be\label{7.1.19}
g(r) = \alpha_{\pm} \sqrt r e^{-\eta_{\pm}r} + ... \, , \ \ 
h (r) = \mp {\alpha_{\pm}\over\sqrt r} e^{-\eta_{\pm}r} + ... \, , 
\ee
\be\label{7.1.21}
f(r) = 1 + \beta e^{-\lambda r} + ...
\ee
where $\alpha_{\pm}$ and $\beta$ are dimension-less constants while the 
dimension-less vector meson mass $\eta_{\pm}$ is given by
\be\label{7.1.22}
\eta_{\pm} = \sqrt{1+{\delta^2\over 4}} \pm {\delta\over 2} \, .
\ee
However, it has subsequently been shown  
that the solution with $\eta_+$ {\it does not} exist for all r. 

On noting that the field Eqs. (\ref{7.1.6}) to (\ref{7.1.8}) are invariant
under $r\rightarrow -r,$ it is easily shown that the behavior of the gauge
and the Higgs fields around r = 0 is given by
\be\label{7.1.23}
g(r) = n +\alpha_1 r^2 +O(r^4) \, , \ \ 
h(r) = \beta + \alpha_1 \delta {r^2\over 2} + O (r^4) \, , 
\ee
\be\label{7.1.25}
f(r) = \alpha_2 r^{\mid n\mid} +O (r^{\mid n\mid +2}) \, .
\ee

Detailed numerical work has subsequently confirmed the existence of the 
radially symmetric
charged vortex solutions with these boundary conditions \cite{7jkkp}.
These correspond to $n$ superimposed vortices. The 
qualitative behaviour of the
charged 
vortex solution which follows from here is as follows : the magnetic 
field
$B$ decreases 
monotonically from its non-zero value at the core of the vortex 
($r=0$)
to reach zero as $r\rightarrow \infty$ with the penetration length 
$1/\eta_-$, while the Higgs
field increases 
from zero at the origin to its vacuum value at infinity with
coherence 
length $1/\lambda$. Finally, the electric field $E_{\rho}$ which 
is radial,
vanishes both at r = 0 
and r = $\infty$ reaching the maximum in between at 
some finite
r. It is worth pointing out that as in the quantum Hall effect, for the 
charged vortex
solutions too, ${\bf E} (\equiv E_{\rho})$ is at right angles to 
${\bf J} (\equiv j_{\theta})$ and
both in turn are at right angles to B.

Why did one obtain two asymptotic solutions for $g$ and $h$, i.e. for the gauge fields 
$A_{\theta}$ and $A_0$?
This is because of the unusual nature of the Higgs 
mechanism
in $2+1$ dimensions in the presence of the CS term. 
Notice that in our case both the Maxwell and the 
CS terms are present and in addition there 
is also Higgs
mechanism in operation. 
Clearly such a theory must still propagate 
only
two massive modes. 
As has been shown in \cite{7pkh},
in this case ${\cal L}_{quad}$ corresponds to Proca 
equation with the CS
term. It propagates a 
self-dual field with two distinct CS type masses and 
that corresponding
to each mass there is one ($P$ and $T$ violating) 
propagating mode. Further, the two masses 
(in dimension-less
form ) are precisely $\eta_{\pm}$ as given by Eq. (\ref{7.1.22}) thereby 
explaining the
reason for the occurrence of two asymptotic solutions $\eta_{\pm}$.

{\bf Vortex-Vortex Interaction}

One of the most interesting question is whether these charged vortices can be
observed experimentally in some planar system. In this context recall that the
neutral (Abrikosov) vortices have been experimentally seen in type-$II$ 
superconductors.
This can be 
understood from the fact that whereas the vortex-vortex interaction
is repulsive in 
the type-$II$ region $(\lambda > 1)$, it is attractive in the 
type-$I$ region
of superconductivity. It is thus of great interest to
study the 
charged vortex-vortex interaction and to see when is it repulsive.
This has been done both in the perturbation theory 
(in the CS mass) and by 
the variational 
calculation \cite{7jkkp}, and in both cases one finds that the charged 
vortex-vortex interaction
is more repulsive than the 
corresponding neutral case with the extra repulsion 
coming from the electric field of the charged vortex. 
For example, when the CS mass is small, then on expanding 
the charged $n$-vortex fields in terms of the corresponding 
neutral vortex 
fields it has been shown that  
\be\label{7.1.35}
E_n (\lambda,\delta)  - n E_1 (\lambda,\delta) = E_n (\lambda, 0) 
- n E_1 (\lambda,0)+ {(n^2-n)\over 4}\delta^2+O(\delta^4)
\ee
so that the 
charged vortex-vortex interaction is always more repulsive than 
the corresponding neutral case.
For example, for 
$\delta = 0.5$, one
finds that the charged vortex-vortex interaction is repulsive even for 
$\lambda > 0.45$ (note
that in the 
neutral case the interaction is repulsive only if $\lambda > 1$).

{\bf Non-Abelian Charged Vortex Solutions}

It is clearly of 
considerable interest to enquire whether the charged vortex 
solutions obtained above can be embedded in 
non-Abelian gauge theories with 
the CS term. The first obvious question 
is whether such vortices could be topologically stable or not.
It is easily seen that if $G$ is the 
gauge group of the non-Abelian gauge theory 
and $H$ is
the sub-group under which the vacuum remains invariant after spontaneous 
symmetry
breaking, then topologically non-trivial vortices are possible only if 
\be\label{7.1.36}
\pi_1 (G/H) \not = 0 \, .
\ee
In the case of $SU(N)$ gauge 
theories, it turns out that no $Z$-vortices are
possible. However, $Z_N$-vortices are possible in case $H$ is 
$Z_N$ since 
$\pi_1(SU(N)/Z_N)=Z_N$. It 
turns out that at least $N$ Higgs multiplets are 
required
so that the vacuum is invariant under $Z_N$ \cite{7dvs}. As a result, 
only one non-trivial charged vortex is possible
in the case of $SU(2)$ gauge theory with 
flux $\Phi = 2\pi/g,$ 
charge $Q = \mu\Phi 
= {2\pi \mu}/g$, 
and angular momentum  
$J = - {Q\Phi/4\pi} = - {\pi\mu/g^2}$ where $g$ is
the gauge coupling constant. 
But since the CS mass 
$\mu$ is 
quantized in non-Abelian
gauge theories having $SU(2)$ as its sub-group i.e.
\be\label{7.1.37}
\mu = {g^2\over 4\pi} n, \ n = 0, \pm 1, \pm 2,... 
\ee
and hence the vortex charge is ${gn/2}$ i.e. it is quantized in units of 
$g/2$
while the angular momentum is quantized in units of $1/4$ i.e.
$J = - {n/4}$.
This is remarkable as 
it strongly suggests that if the usual spin-statistics
connection is 
valid then whereas the Abelian charged vortex is an anyon with 
any phase
factor, the non-Abelian ($SU(2)$) charged vortex can only be a semion, 
a fermion or a boson.

{\bf Relativistic Pure Chern-Simons Vortices}

We have obtained above the charged vortex solutions in case the gauge
part of the Abelian Higgs model consists of both the Maxwell and the 
CS term. It may be
of some interest to enquire 
whether the Abelian Higgs model with pure CS 
term can also
admit charged vortex solutions. This question is 
specially relevant in the 
context of
condensed matter systems since in the long wave length limit, the CS 
term having one derivative dominates over
the Maxwell term which has two derivatives. It turns out that the 
answer to the 
question is yes 
\cite{7jk}.

In the 
absence of the Maxwell term and with the same rotationally symmetric 
ansatz
as in 
Eq. (\ref{7.1.3}), it follows from Eqs. (\ref{7.1.6}) and (\ref{7.1.7}) 
that the
gauge field equations are already of first order. 
However, Eq. (\ref{7.1.8}), for the Higgs fields, is still
a coupled second 
order equation. We now show that in case one replaces the 
standard
double well $\phi^4$-type 
potential by the following $\phi^6$-type potential \cite{7hkp} 
\be\label{7.2.3}
V(\mid\phi\mid) = 
{e^4\over 8\mu^2} \mid\phi\mid^2 (\mid\phi\mid^2-C_0^2)^2
\ee
then even the Higgs field satisfies a first order equation.
It is worth pointing out here that whereas a Higgs potential of the type 
$\sum_i C_i\mid\phi\mid^i$ with
$0\leq i \leq 4$ is renormalizable in $3+1$ dimensions, 
$\sum_i C_i\mid\phi\mid^i$ with $0\leq i \leq 6$ is renormalizable in $2+1$ 
dimensions.

When the Maxwell term is absent and the Higgs potential is as given by 
(\ref{7.2.3}), the 
vortex energy (\ref{7.1.9}) 
can be rewritten as
\be\label{7.2.7}
E_n = \pi C_0^2 \int^{\infty}_0 r d r \bigg [ (f'\mp {1\over r} fg)^2 + 
f^2 [h\mp {(1-f^2)\over 2\delta}]^2\mp {1\over r} {d\over dr}
[(1-f^2)g]\bigg ] .
\ee
This gives a rigorous lower bound on the energy in terms of the flux
\be\label{7.2.8}
E_n \geq \pm \pi C_0^2 [g(0)-g(\infty) ] \equiv \pm {1 \over 2} e C_0^2\Phi 
\ee
since the 
finite energy consideration requires that $f^2 g$ vanish at both the ends. 
This
bound is saturated when the following self-dual first order equations are 
satisfied
\be\label{7.2.9}
f'(r) = \pm {1\over r} fg
\ee
\be\label{7.2.10}
-{1\over r} g'(r) = {hf^2 \over \delta} = \pm {1\over 2\delta^2} f^2(1-f^2) .
\ee
It is easily checked that these first order equations are consistent with the 
second order field Eq. (\ref{7.1.8}). One
can in fact decouple these coupled first order equations and show that the
Higgs field $f$ must satisfy the following un-coupled second order equation
\be\label{7.2.11}
f''(r) + {1\over r} f'(r) - {f'^2(r) \over f} +{1\over 2\delta^2} f^3(1-f^2) = 0 \, .
\ee
Several comments are in order at this stage.
\begin{enumerate}
\item These self-dual equations are similar to those of the
Nielsen-Olesen (neutral) self-dual vortices 
(which are valid only if $\lambda = 1$).
\item Whereas the Lagrangian for the self-dual neutral vortex case (i.e.
Lagrangian (\ref{7.1.1}) with $\mu = 0$ and $\lambda =1$) is the bosonic
part of a N = 1 supersymmetric theory \cite{7dvf},
the Lagrangian for the self-dual charged vortex case (i.e. the Lagrangian 
(\ref{7.1.1}) with the Maxwell
term being absent and the Higgs potential being as given by Eq. (\ref{7.2.3})) 
is the
bosonic part of a N = 2 supersymmetric theory \cite{7lee}.
\item The $\phi^4$-potential as given in 
Eq. (\ref{7.1.1}) and the $\phi^6$-
potential as given by Eq. (\ref{7.2.3}) represent very different physical 
situations.
For example, whereas the $\phi^4$-potential 
corresponds to the case of the 
second order phase
transition with $T < T_c^{II}$, the $\phi^6$-potential as given in 
Eq. (\ref{7.2.3}) corresponds
to the case of first order phase transition with $T = T_c^I$ \cite{7bk}. 
\item The nature of 
Higgs mechanism when only Chern-Simons term is present is
somewhat unusual \cite{7dy}.
One finds that in the 
limit $e^2 \rightarrow \infty, \ \mu \rightarrow \infty$,
with their ratio fixed,
the mass $m_+$ decouples from the theory.  
Thus in the case of the pure CS term, one finds that 
after the Higgs
mechanism, the gauge field is massive and propagates 
one mode. 
\end{enumerate}

Let us now discuss the most remarkable property of the self-dual Eqs.  
(\ref{7.2.9}) and (\ref{7.2.10}).
In particular, since the Higgs potential (\ref{7.2.3}) has degenerate minima at 
$\mid\phi\mid=0$ and $\mid\phi\mid = C_0$, hence, 
it turns out that at the 
self-dual 
point, one can simultaneously have both the topological and the non-topological charged
vortex solutions. It is worth pointing out that at the time of this 
discovery, 
no other self-dual system was known which exhibited this
remarkable property. 

{\bf Topological Self-dual Solutions }

The topological, self dual charged vortex solutions satisfy the same
boundary conditions as given by Eqs. (\ref{7.1.10}) and (\ref{7.1.11}) 
with $\beta\equiv h (r=0)
= \pm {1/2\delta^2}$. Note that the upper (lower) sign 
corresponds to
$ n > 0 (< 0)$. As a result, the flux $\Phi$, the Noether charge $Q$, and 
the angular 
momentum $J$ of these charged vortices are again as given by 
Eqs. (\ref{7.1.12}), (\ref{7.1.14}) and 
(\ref{7.1.17}) respectively while the energy of these charged vortices is 
$\pi C^2_0 \mid n \mid.$ 
From now onwards, we shall confine our discussion to the case of 
$n > 0$ i.e.
those corresponding to the upper choice of sign. Solution with 
$n < 0$ are related to these by
the transformation $g\rightarrow -g, \ f\rightarrow f$. 

A countable infinite number of sum rules have been derived 
\cite{7k} and using
the first two, it has been proved that the magnetic moment 
of the topological, self-dual charged
$n$-vortex is given by \cite{7kh} 
\be\label{7.2.14}
K_z = 2\pi n (n+1) {\delta^2\over e} \, .
\ee
Note that for the neutral $n$-vortex, $K_z = \Phi = {2\pi n/e}$. 

No analytic topological 
self dual charged vortex solution has been 
obtained as yet. However,
one can show that all the fields approach their asymptotic values 
exponentially
fast. 
It may be  note that at the 
self-dual point, the vector and the
scalar meson masses are equal.  
Further, whereas
for the Maxwell-CS case, the magnetic field is maximum at the core of 
the vortex
($r\rightarrow 0)$, for the 
pure CS vortices, the magnetic field is zero at 
the core
of the vortex and is concentrated in a ring surrounding the vortex 
core.

{\bf Non-topological Self-dual Solutions}

Since $\mid\phi\mid = 0$ as well as $\mid\phi\mid = C_0$ are 
degenerate minima
of the Higgs potential (\ref{7.2.3}), hence it turns out that one 
could also have 
non-topological self-dual \index{charged vortex solutions!non-topological 
self-dual} charged vortex solutions \cite{7kh,7jlw}. In this case, the 
finite energy considerations demand the following boundary conditions
\be\label{7.2.19}
\lim_{r\rightarrow\infty} \ \ f(r) = 0 \, , \ 
g(r) = \mp \alpha \, , \ \alpha > 0
\ee
\be\label{7.2.20}
\lim_{r\rightarrow 0} \ \ f(r) = 0 \, , \ g(r) = n \ \ \mathrm{for} \ n \not = 0
\ee
\be\label{7.2.21}
\lim_{r\rightarrow 0} \ \ f(r) = \eta \, , \  g(r) = 0 \ \ \mathrm{for} \ n  = 0
\ee
where $\eta$ is an arbitrary number while 
$-\alpha (+\alpha)$ is for $n > 0 ( < 0)$. As a result, the flux, the charge, 
the energy
and the angular momentum of these vortices for $(n > 0)$ are   
\bea\label{7.2.22}
 & \Phi & = {2\pi\over e} (n+\alpha) \, , \ Q = \mu\Phi 
= {2\pi\mu\over e} (n+\alpha) \, , \nonumber\\
 & J & = {\pi\mu\over e^2}
(\alpha^2-n^2) \, , \ E = \pi C^2_0 (n+\alpha) \, .
\eea
 Note that unlike the topological case, the angular momentum is no 
more equal to 
$-{{Q\phi}/4\pi}$. Here $\alpha$ is a positive number but how much is it? The 
finiteness of energy
requires that $\alpha > 1$ but otherwise $\alpha$ seems to be 
completely arbitrary. However,
it is not so and we now show \cite{7kha} that $\alpha$ satisfies 
a rigorous lower 
bound of $\alpha \geq n +2$. To this end, consider the self-dual 
Eq. (\ref{7.2.10}). On
 integrating both sides of this equation and using boundary 
conditions (\ref{7.2.19}) to
(\ref{7.2.21}), one obtains (for $n > 0$)
\be\label{7.2.23}
-\int^{\infty}_0 {dg\over dr} dr = n+\alpha 
= {1\over 2\delta^2} \int^{\infty}_0 r dr f^2(1 - f^2) > 0 \, .
\ee
Similarly, on using Eqs. (\ref{7.2.9}) and (\ref{7.2.10}) we 
have on integration
\be\label{7.2.24}
\int^{\infty}_0 g {dg\over dr} dr =  {1\over 2} (\alpha^2-n^2) 
= -{1\over 2\delta^2} \int^{\infty}_0 r^2 f(1-f^2) {df\over
dr} dr \, .
\ee
On integrating by parts and using the fact that $r^2f^2$ and $r^2f^4$ 
vanish as 
$r\rightarrow\infty$ (note $f(r) \sim r^{-\alpha}$ with $\alpha > 1 $ as 
$r\rightarrow\infty$), we then have
\be\label{7.2.25}
(\alpha^2-n^2) = {1\over \delta^2}\int^{\infty}_0 r dr 
(f^2 - {1\over 2} f^4) \, .
\ee
On combining the two sum rules, we then have 
\be\label{7.2.26}
(\alpha+n) (\alpha-n-2) = {1\over 2\delta^2} \int^{\infty}_0 r dr f^4 \geq 0
\ee
which gives us a rigorous lower bound on $\alpha$ i.e.
$\alpha \ \geq \ n+2$.
It turns out that this bound is 
never saturated in the relativistic case. However, as 
we shall see below, it is indeed saturated in the case of the 
non-relativistic 
self-dual non-topological charged vortices. It may be noted here that 
there is however no upper bound on 
$\alpha$. We thus conclude
that the flux of 
the relativistic non-topological vortices must necessarily be 
greater than
${{4\pi (n+1)}/e}$. More remarkable is the fact that whereas the 
angular 
momentum of the topological vortices is always negative and 
proportional to $n^2$, the angular
momentum of the non-topological vortices, on the other hand, is 
necessarily positive
and in general is not proportional to $n^2$. Further, the magnetic 
moment of
the non-topological vortices has also been computed analytically by 
using the sum rules and
shown to be negative \cite{7kha} 
\be\label{7.2.28}
K_z = - {2\pi\delta^2\over e} (\alpha+n) (\alpha-n-1) < 0 \, .
\ee
Note that the magnetic moment of the topological vortices is on the 
other hand
always positive.

Are these non-topological vortices stable or do they decay to the charged scalar
meson ? This question has been discussed \cite{7jkh} and it
has been shown that as far as the decay to the scalar meson is concerned, 
these
non-topological solitons are at the edge of their stability. In particular, 
using $E$ and 
$Q$ as given by Eq. ({\ref{7.2.22}) and 
noting that the mass $m$ of the scalar particle
in the symmetric 
vacuum is $e^2c^2_0/2\mu$, it follows that $E = m Q/e$. Thus the 
stability does not 
impose any upper bound on the charge of the non-topological soliton.
No analytic solutions of Eqs. ({\ref{7.2.9}) and 
(\ref{7.2.10}) have been
obtained as yet in the non-topological self-dual case. 
However, the behavior 
of the fields near $r\rightarrow 0$ and for large $r$ is 
easily obtained. 
In particular, using the boundary conditions (\ref{7.2.19}) to 
(\ref{7.2.21}), 
it is not difficult to 
show that for $r\rightarrow\infty$, the $n = 0$ vortex solution
has the behavior
\be\label{7.2.29}
g(r) = - \alpha + {G_0^2\over 4(\alpha-1)(r/\delta)^{2\alpha-2}} 
+ O((r/\delta)^{-4\alpha+4})
\ee
\be\label{7.2.30}
f(r) = {G_0\over (r/\delta)^{\alpha}} 
- {G_0^3\over 8(\alpha-1)^2(r/\delta)^{3\alpha-2}} 
+ O((r/\delta)^{-5\alpha+4}) \, .
\ee
On the other hand, as $r \rightarrow 0$, while $f(0)$ is not constrained, 
$g(0)$ must vanish so as to have a non-singular solution.  
Thus for the $n = 0$ non-topological vortex, the magnetic field  
(${-g'(r)/r}$) is 
maximum at the core of the vortex ($r = 0$) and falls off with 
a power law fall 
off as $r \longrightarrow \infty$. Note, however, that the magnetic 
field for the topological CS vortices is zero at the core, 
and is maximum in a ring surrounding the core of the vortex.

Finally, let us 
consider the behavior of the $n \neq 0$ (we as usual consider $n > 0)$ 
non-topological
self-dual charged vortex solutions. It is easily shown that these 
solutions are 
hybrids of the two previous cases i.e. their large distance behavior 
is the same as
those of the $n = 0$ non-topological charged vortex solutions as given by 
Eqs. (\ref{7.2.29}) and (\ref{7.2.30}). On the 
other hand their short distance behavior is the same as
those of the self-dual topological charged vortex solutions. 
Thus for $n \neq 0$ non-topological vortices, 
the magnetic field
vanishes at the core of the vortex and falls off with a power law fall 
off as $r\rightarrow\infty$.

It is worth pointing out that since the $\phi^6$-potential as given by 
Eq. (\ref{7.2.3}) has
two disconnected but degenerate vacua at $\mid\phi\mid = 0$ and 
$\mid\phi\mid = C_0$, 
hence, apart from the charged vortex solutions, they also possess one
dimensional domain wall solutions \cite{7jlw,7bk}.

So far, we have only discussed the self-dual rotationally symmetric 
CS vortices.
However, the self-dual solutions can in fact be obtained even without choosing 
the
rotationally symmetric $n$-vortex {\it ansatz} (\ref{7.1.3}). 
Further, rigorous arguments have subsequently been given  
for the existence of the self-dual
topological \cite{7wa} and non-topological \cite{7sy} charged vortex solutions 
even when the vortices 
are not 
superimposed on each other but lie at arbitrary positions in the plane. Let us
note an interesting fact about the angular momentum of these charged vortices.
For example, whereas the angular momentum of the $n$ superimposed topological 
vortices is 
$n^2$ times that of a single vortex, the angular momentum of the $n$ topological
vortices (each of which has unit vorticity) which are well separated from 
each other,
is only $n$ times the angular momentum of the single vortex. However, the 
energy,
flux and the 
charge of the $n$ vortices in both the cases is the {\it same}. 
Thus we see that whereas the energy, flux and charge, are 
the global quantities, the angular momentum
of a configuration depends on the local behavior. 

A zero-mode analysis of the spectrum of small fluctuations \cite{7jlw}
around the self-dual vortices indicates that 
whereas the number of zero modes in the case of the topological
self-dual vortices is $2n$, in the non-topological 
case, the same number is $2n+2[\alpha]$ where $[\alpha]$ 
denotes the integer part
of $\alpha$. In the topological case, this number is identified with the 
number of
parameters required to 
describe the location of the $n$ vortices while the 
counting is
less clear in the non-topological case. 

{\bf Interaction Between Self-Dual CS Vortices}

The slow motion of the Abelian self-dual CS 
vortices has been analyzed \cite{7kmi} using
Manton's technique \cite{7man}. In this approach, 
one constructs an effective 
quantum mechanical Lagrangian (not density) which 
describes the fluctuations about the 
static self-dual classical configurations and not surprisingly,
one obtains a statistical interaction term. Further one also obtains
a term corresponding to the velocity dependent Magnus force. It
turns out that this force is in fact necessary in order to have 
correct spin-statistics relation. 

Self-dual charged vortices have also been  
obtained in the original $\phi^4$-type model itself 
by adding a neutral scalar field to Eq. (\ref{7.1.1}) and 
changing the
$\phi^4$-potential suitably \cite{7lm}. 

Finally, semi-local self-dual CS vortices have been obtained in an Abelian
Higgs model with pure CS term \cite{7khare} and with 
$SU(N)_{global}\otimes U(1)_{local}$ symmetry. 
The interesting 
point is that the semi-local vortices, even though topologically
trivial, are 
stable under small perturbations due to the gradient energy term.

{\bf Non-relativistic Chern-Simons Vortices}

Let us now discuss the 
non-relativistic  
limit of the Abelian Higgs model with the pure
CS term. 
The Lagrangian density for the Abelian Higgs model with pure CS 
term is given by 
\be\label{7.3.1}
{\cal L} = {1\over 2} (D_{\mu}\phi)^* (D^{\mu}\phi)+{\mu\over 4} \varepsilon_{\mu\nu\lambda} 
F^{\mu\nu} A^{\lambda} - {e^4\over 8c^4\mu^2} \mid\phi\mid^2 (\mid\phi\mid^2 
- C_0^2)^2 \, ,
\ee
where the Higgs potential is as given by Eq. (\ref{7.2.3}). Here 
we write all 
the
factors of the velocity of light $c$ explicitly since we are 
considering the 
non-relativistic
limit of a relativistic theory.
Let us first note that the quadratic term in the Higgs potential defines 
the mass through
its 
coefficient ${m^2c^2/ 2}$. Comparison with Eq. (\ref{7.3.1}) shows that
$C_0^2$ must have the value
$C^2_0 = (2\mid\mu\mid m c^3)/e^2$
so that the Lagrangian density (\ref{7.3.1}) can be rewritten as
\bea\label{7.3.3}
{\cal L} & = & {1\over 2 c^2} \mid (\partial_t-{ie\over\hbar}A^0) \phi \mid^2 
- {1\over 2}
\mid{\bf D}\phi\mid^2 - {m^2c^2\over 2} \mid\phi\mid^2 \nonumber\\ 
         & + & {m e^2\over 2c\mid\mu\mid}\mid\phi\mid^4 
- {e^4\over 8c^4\mu^2} \mid\phi\mid^6 
+{\mu\over 4} \varepsilon_{\mu\nu\lambda} F^{\mu\nu} A^{\lambda} \, .
\eea
The non-relativistic limit $(c\rightarrow\infty)$ now proceeds in the 
standard manner.
On writing the mode expansion of the scalar field $\phi$ as 
\be\label{7.3.4}
\phi = {1\over\sqrt m} \bigg [ e^{-imc^2t} \ \psi 
+ e^{imc^2t} \bar\psi^* \bigg ]
\ee
and substituting it in Eq. (\ref{7.3.3}), dropping all terms that either 
oscillate
as $c\rightarrow\infty$ or are sub-leading in powers of c, the matter 
part of
the Lagrangian density can be shown to be
\be\label{7.3.5}
{\cal L} = i \psi^* D_0\psi - {1\over 2m} \mid {\bf D} \psi \mid^2 
+{e^2\over 2mc\mid\mu\mid}
\rho^2 +{\mu\over 4} \varepsilon_{\mu\nu\lambda} F^{\mu\nu} A^{\lambda} \, .
\ee
Here $\rho = \psi^* \psi$ is the matter density of particles and we have dropped
the anti-particle 
part from the Lagrangian density (i.e. we are working in the zero 
anti-particle
sector) by setting $\bar\psi = 0$ 
since the particle and the anti-particle 
parts
are separately conserved. The remarkable fact is that one now has an {\it 
attractive}
quartic $(\rho^2)$ self-interaction. This non-relativistic model can be 
looked upon
either as a non-relativistic classical field theory or as a second quantized 
$N$-body
problem with $2$-body attractive delta-function interaction.

The Euler-Lagrangian equations of motion which follow from the Lagrangian 
density
(\ref {7.3.5}) are
\be\label{7.3.6}
-{1\over 2m} {\bf D}^2 \ \psi - {e^2\over mc\mid\mu\mid} \mid\psi^2\mid \psi 
- i D_0 \psi = 0
\ee
\be\label{7.3.7}
F_{\mu\nu} = - {1\over\mu} \varepsilon_{\mu\nu\rho} J^{\rho}
\ee
where  $J^{\mu}\equiv (\rho, \vec J)$ is a Lorentz covariant notation for the 
conserved non-relativistic charge and current densities i.e.
\be\label{7.3.8}
\rho = \mid \psi^2 \mid , \ J^k = - {i\hbar^2\over 2m} [\psi^* D^k \psi 
- (D^k \psi)^*
\psi ] \, .
\ee
The field Eqs. 
 (\ref{7.3.6}) and (\ref{7.3.7}) are together termed as the {\it planar 
gauged nonlinear
Schr\"{o}dinger equations}. The gauge field Eq. (\ref{7.3.7}) can also be 
re-expressed as
\be\label{7.3.9}
B \equiv F_{12} = {e\over\mu} \rho
\ee
\be\label{7.3.10}
E^i \equiv F_{i0} = - {e\over c\mu} \varepsilon^{ik} J_k \, . 
\ee
From here, we immediately obtain the fundamental relation between the Noether 
charge Q and 
the magnetic flux $\Phi$ i.e. $Q = \mu\Phi$. As in the relativistic case, it
is easily checked that the second order field Eqs. (\ref{7.3.6}) and 
(\ref{7.3.7}) are
solved by Eq. (\ref{7.3.9}) and the self-dual ansatz 
\be\label{7.3.11}
D_j \psi = \pm i \varepsilon_{jk} D_k \psi
\ee
in the case of the static 
solutions with $A_0$ chosen as
\be\label{7.3.12}
A_0 = \mp {e\over 2m\mu c}\mid\psi\mid^2 \, .
\ee
Here we have made use of the following factorization identity
\be\label{7.3.14}
{\bf D}^2 \psi = D_{\pm} D_{\mp} \psi \mp {e\over c} F_{12} \psi \, .
\ee

We now show that the self-dual Eqs. (\ref{7.3.9}) and (\ref{7.3.11}) can be 
solved
completely and explicitly. On writing the complex field $\psi$ as
$\psi = e^{-i\omega} \rho^{1/2}$ 
the self-duality Eq. (\ref{7.3.11}) yields the vector potential 
\be\label{7.3.16}
A_i = \partial_i \omega \pm {c\over 2e} \varepsilon^{ij}\partial_j \ln \rho
\ee
which is valid away from the zeros of $\rho$. On inserting this form of 
${\bf A}$ into
the other self-dual Eq. (\ref{7.3.9}) yields the
famous Liouville equation
\be\label{7.3.17}
{\bf \nabla}^2 \ln \rho = - {2e^2\over c\mid\mu\mid} \rho
\ee
which is known to be integrable and completely solvable and which must be 
solved away
from the zeros of $\rho$. It is worth noting that with our sign conventions, we
have the 
Liouville equation with the correct sign in that only such an equation has
real, positive, regular solutions. The most general such solution is known to be
given by
\be\label{7.3.18}
\rho ={c\mid\mu\mid \mid f'(z) \mid^2\over e^2[1+\mid f(z)\mid^2]}
\ee
where $f(z)$ is any holomorphic function and $z = r e^{i\theta}$.  
Explicit radially symmetric solutions may be obtained by taking
$f (z) = ({z/z_0})^{\pm n}$.
The corresponding self-dual charge density is
\be\label{7.3.20}
\rho = {4\mid\mu\mid n^2 c\over e^2 r^2_0} {(r/r_0)^{2(n-1)}\over [1+(r/r_0)^{2n}]^2}
\ee
which behaves like $r^{2(n-1)}$ as $r\rightarrow 0$ while as 
$r\rightarrow\infty$, it behaves 
like $r^{-2-2n}$. Thus $\rho$ is regular at the origin if $n \geq 1$. From
Eq. (\ref{7.3.16}) it then follows that as $r\rightarrow 0$, the vector 
potential
behaves as
\be\label{7.3.21}
A_i (r) \sim \partial_i \omega \pm {c(n-1) \over e} \varepsilon_{ij} {x^j\over r^2}
\ee
i.e. it is singular at $r = 0$. This singularity is removed if we choose
$\omega = \pm {{c (n-1)\theta}/e}$. Thus the profile of the self-dual 
$\psi$ field
is given by
\be\label{7.3.22}
\psi ({\bf r}) ={2n\sqrt{\mid\mu\mid c}\over e r_0} 
{(r/r_0)^{n-1}\over [1+(r/r_0)^{2n}]}
\ e^{\pm i(n-1)\theta} .
\ee
On requiring that $\psi$ be single valued, we then find that $n$ must be an 
integer, and for $\rho$
to have decaying behavior as $r\rightarrow\infty$, we require that $n$ must be 
positive.

Several comments are in order at this stage.
\begin{enumerate}
\item  Integrating $\rho$ as given in (\ref{7.3.20}) over all space yields $n$ 
(the total number of particles) and hence the flux 
(in view of Eq. (\ref{7.3.9})). We obtain
$\Phi = {(4\pi cn/e)}$ with $n$ = 1,2,...
which means that this configuration carries an even number of flux units. 
This is
in contrast to the relativistic case where the flux unit need not necessarily 
be even. Further, note that unlike the relativistic non-topological case, here 
the lower bound on $\alpha (\ge n+2$) is saturated. 
As has been shown \cite{7kim}, this is because of 
the special inversion symmetry of the Liouville equation. In particular, notice
that the Liouville equation is invariant under the transformations 
\be\label{7.3.23a}
r\rightarrow 1/r, \ \theta
\rightarrow\theta, \ \rho(r) \rightarrow\rho (1/r) = r^4 \rho (r) \, . 
\ee
As a result, the 
behavior of $\rho$ at infinity is uniquely determined by its behavior at  
the origin
thereby fixing $\alpha = n +2$.  
\item It is worth pointing out the $Q, \Phi$ and $J$ for the non-relativistic 
charged vortices
are the same as those for the relativistic non-topological charged vortices as 
given by 
Eq. (\ref{7.2.22}) provided one chooses $\alpha = n+2$ (note that in the 
non-relativistic
case, $n = 1,2,...$ while n = 0,1,2,... in the relativistic case).
\item  The radially symmetric solution (\ref{7.3.22}) was obtained by choosing 
the
holomorphic function $f(z) \propto (z)^{-n}$ and corresponds to $n$ solitons 
superimposed at the origin with common scale factor $r_0$. The most general
solution
corresponding to $n$ separated solitons may be obtained by taking
\be\label{7.3.25}
f (z) = \sum^n_{i=1} {\alpha_i\over (z - z_i)}
\ee
where $2n$ real parameters $z_i$ describe the location of the solitons and $2n$ 
real
parameters $\alpha_i$ correspond to the scales and the phases of 
the solitons. Thus the solution depends on $4n$ parameters. Using an index 
theory calculation \cite{7ksye} it has been shown that this is the most general 
solution.
\end{enumerate}

\section{$CP^1$ Solitons With Hopf Term}

In this section we discuss the extended (neutral) anyon solutions in 
relativistic field theories. Historically, such solutions were first
written down in the case of $O(3)$   
$\sigma$-model with Hopf
term in 2+1 dimensions \cite{7wz}. 
Unfortunately, in this case, the Hopf
term cannot be written down as a local function of the basic fields of the 
theory.
Therefore, we shall discuss the essentially equivalent example of the $CP^1$ 
model with
the Hopf term since in this case the Hopf term can be written down as a local
function of the basic fields of the theory \cite{7wuz}.

The action for the $CP^{1}$ model in 2+1 dimensions
is given by
\be\label{7.4.1}
I = \int d^3 x \ (D_{\mu} z)^* (D^{\mu} z)
\ee
where $D_{\mu} z \equiv (\partial_{\mu}-iA_{\mu})z $ with $z 
= (z_1,z_2)$ being
a complex vector fulfilling $\mid z\mid^2 = 1$. Note that $A_{\mu}$ here 
does not
represent independent degrees of freedom, but is entirely determined in 
terms of $z(x)$
through the constraint equation
\be\label{7.4.2}
A_{\mu} = - i z^* \ \partial_{\mu} z \, .
\ee
The action (\ref{7.4.1}) is invariant under the local $U(1)$ 
transformations
\be\label{7.4.3}
z_a (x) \rightarrow z_a (x) e^{i\Lambda (x)}, 
\ A_{\mu} (x)\rightarrow A_{\mu} (x)
+\partial_{\mu} \Lambda (x) \, .
\ee

As is well known, the $CP^1$ model admits self-dual 
soliton solutions. To 
obtain them,
let us first note that the field equation is obtained by extremizing the 
action
(\ref{7.4.1}) with respect to $z(x)$ subject to the constraint $\mid z\mid^2 
= 1$. This
constraint is best introduced in the variational formalism by using a 
Lagrangian
multiplier i.e. one extremizes $I+\int d^3x \lambda (x) (z^* z-1)$.
The resulting field equation is
\be\label{7.4.4}
(D_{\mu}D^{\mu}+\lambda) z = 0 \, .
\ee
The Lagrange multiplier $\lambda(x)$ is eliminated by using 
$\lambda = \lambda z^* z = - z^* D_{\mu} D^{\mu} z$.
Let us now consider the static solutions. In this case, the 
field equation (\ref{7.4.4}) reduces to 
\be\label{7.4.6}
{\bf \nabla}^2 z - (z^* \cdot {\bf \nabla}^2 z) z = 0 \, .
\ee
The energy of a static solution as obtained from the action (\ref{7.4.1}) 
is clearly
\be\label{7.4.7}
E = \int (D_i z)^* (D_i z)  d^2 x \, , \ \ i = 1,2 \, .
\ee
Finiteness of energy requires 
that as $r\equiv \mid {\bf x} \mid \rightarrow\infty$, 
$D_i z \equiv \partial_i z - i A_i z = 0$.

Let us start from the topological inequality which follows from
\be\label{7.4.9}
\bigg [ (D_i z)^* \pm i \varepsilon_{ij} (D_j z)^* \bigg ] \cdot  
\bigg [D_i z \mp i \varepsilon_{ik} D_k z \bigg ] \geq 0 \, .
\ee
Because of the constraint $\mid z\mid^2=1$, this inequality can be 
re-expressed in the form
\be\label{7.4.10}
(D_iz)^*\cdot (D_i z) \geq \varepsilon_{ij} (D_i z)^*\cdot (D_j z)
\ee
so that the energy is bounded from below by the topological charge $Q$
i.e. $E \geq 2\pi \mid Q\mid$,
where
\be\label{7.4.12}
Q = -{i\over 2\pi} \int d^2 x \ \varepsilon_{ij} (D_i z)^* \cdot (D_j z) \, .
\ee
In any $Q$-sector, the energy reaches its minimum when the fields minimize 
the
energy in that sector and satisfy the first order self dual field equation
\be\label{7.4.13}
D_i z = \pm i \ \varepsilon_{ij} D_j z \, .
\ee
Note that the solutions of Eq. (\ref{7.4.13}) automatically solve the second
order field Eq. (\ref{7.4.6}) 
while the converse need not be true. 

The most general solution 
for $z$ can be written down in terms of (anti) 
holomorphic
function $\omega$
\be\label{7.4.14}
z = {1\over\sqrt{1+\mid \omega\mid^2}} \bigg (\matrix{ \omega\cr 1\cr}\bigg ) .
\ee
These solutions are characterized by the energy $E = 2\pi \mid Q\mid$ where $Q$ 
is as
given by Eq. (\ref{7.4.12}). One can in fact define a topological current 
$J^{\mu}$
\be\label{7.4.15}
J^{\mu} = -{i\over 2\pi} \varepsilon^{\mu\nu\lambda} 
(D_{\nu} z)^* (D_{\lambda}z)
\ee
which is conserved by construction, and the topological charge $Q$  
as given above, is related to it  
by $Q = \int J^0 d^2 x$. One can easily show that for the soliton solutions, 
$Q$ is
just the winding number i.e. $Q$ clearly describes the homotopy of the mapping
${\bf S}_2\rightarrow{\bf S}_2$.

Since $J^{\mu}$ is the topological conserved current, hence one can clearly add
the following gauge invariant action 
\be\label{7.4.16}
I_H = \int d^3 x {\theta\over 2\pi} A_{\mu}  J^{\mu}
\ee
to the original action (\ref{7.4.1}) .
This action is nothing but the Hopf term which 
is related formally to the CS term since from 
Eqs. (\ref{7.4.2}) and (\ref{7.4.15}) it follows that
\be\label{7.4.17}
A_{\mu} J^{\mu} = {1\over 4\pi} \varepsilon^{\mu\nu\lambda} 
A_{\mu} F_{\nu\lambda} \, .
\ee
Note however that here $A_{\mu}$ is not an independent gauge field but is 
entirely determined in terms of $z(x)$ 
through the constraint Eq. (\ref{7.4.2}). 
As a result,
unlike the CS term, the 
Hopf term is locally a total divergence and hence 
does not
contribute to the equations of motion. 

Note that unlike the CS term, the Hopf 
term has no dynamics. Besides, for the $CP^1$ soliton solutions
(which are time independent solutions of the equations of motion), the Hopf
term is identically zero because of the time derivative and the relationship
(\ref{7.4.2}). Thus the way the Hopf term imparts fractional spin and 
statistics to
the soliton is similar to that in quantum mechanics but it is very 
different than the way the CS term imparts fractional spin and 
statistics. In particular,
since the Hopf density is a total divergence, hence the Hopf action can be 
expressed in terms of the surface terms, namely 
two integrals at the initial and 
final times
so that in the path integral formalism, the contribution of this action is 
essentially
in terms of the phases of the initial and the final wave functions. Since the
configuration space in question is multi-connected, the Hopf action depends 
on the
homotopy classes of the path and, therefore, the converted phases are 
multi-valued
which in turn gives rise to the fractional spin ($= \theta /2\pi$) and the 
solitons obey fractional statistics characterized by $\theta$ \cite{7wz,7wuz}.

\section{Anyons as Elementary field Quanta}

In this section we enquire whether one can
construct local quantum field theories where the fundamental fields 
represent the creation and annihilation of anyons.
Let us consider a complex bosonic non-relativistic matter field 
$\psi({\bf x},t)$ of mass
$m$ (of course a similar discussion can also be done for the fermionic matter 
field).
Let us minimally couple it to an Abelian gauge field $A_{\mu}$ with a 
CS kinetic term \cite{ak,8le}
\be\label{8.1}
S =  \int d^3 x [ i\psi^+D_0\psi
+{1\over 2m}\psi^+(D^2_1+D^2_2)\psi
+{\mu\over 2}\varepsilon^{\mu\nu\lambda} A_{\mu}\partial_{\nu}A_{\lambda} ]
\ee
where $D_{\mu}=\partial_{\mu}-ie A_{\mu}$ is the covariant derivative. For
simplicity, in this section we shall set $\hbar = c = 1$. 
On varying the action with respect to $A_{\mu}$, we obtain 
\be\label{8.2}
\varepsilon^{\mu\nu\lambda} F_{\nu\lambda} = {2e\over\mu} J^{\mu}
\ee
where the current $J^{\mu}$ is explicitly given by
\be\label{8.3}
\rho \equiv J^0 = \psi^+\psi \, , \ \ 
J^k = {1\over 2mi}[\psi^+ D^k\psi-(D^k\psi)^+\psi] \, .
\ee
Here $\rho$  and ${\bf J}$ are the number density and the current density 
operators respectively which satisfy the continuity equation
$\partial_t \rho +{\bf \nabla}.{\bf J} = 0$.
As seen in previous sections, Eq. (\ref{8.2}) is a remarkable relation
indicating that the CS field strength is completely determined by
the particle current. Even more remarkable is the fact that the gauge potential
$A_{\mu}$ itself is not an independent degree of freedom.

Let us consider the $\mu = 0$ component of Eq. (\ref{8.2})
\be\label{8.6}
B = {e\over \mu}\rho
\ee
where $B = {\bf \nabla} \times {\bf A}$ is the CS magnetic field. 
This equation
is clearly the second quantized version of the Gauss law constraint obtained 
in the last two chapters (except that whereas in those cases $\rho$ was the 
charge density, here $\rho$ is the
matter density, hence the extra factor of $e$ in Eq. (\ref{8.6}) compared to 
those cases). Now, in the weyl 
gauge
$\partial_i A^{i}=0$. Hence, 
one can invert Eq. (\ref{8.6}) without any ambiguity 
and solve
for the vector potential ${\bf A}$. We obtain
\be\label{8.7}
A^i(x) =\varepsilon^{ij}{\partial\over\partial x^j}\bigg ({e\over\mu}\int d^2 y 
G({\bf x}-{\bf y})\rho(y)\bigg )
\ee
where $G$ is the two-dimensional Green function
\be\label{8.8}
{\bf \nabla}^2 G({\bf x } - {\bf y}) =\delta ({\bf x} - {\bf y})
\ee
whose solution is well known to be
\be\label{8.9}
G({\bf x} - {\bf y}) = {1\over 2\pi} \ln 
(p\mid {\bf x}-{\bf y} \mid)
\ee
where $p$ is an arbitrary scale. Thus $A^i$ can be written as
\bea\label{8.10}
A^i(x) & = & \varepsilon^{ij} {\partial\over\partial x^j}
\bigg [{e\over 2\pi\mu}\int d^2 y
\ln \mid {\bf x}-{\bf y} \mid \rho(y) \bigg ] \nonumber\\
       & = & - {e\over 2\pi\mu} \int d^2 y
{\partial\over\partial x^i}\phi({\bf x} - {\bf y}) \rho(y)
\eea
where $\phi$ is the winding (polar) angle i.e.
\be\label{8.11}
\phi({\bf x} - {\bf y})= \arctan ({x^2-y^2\over x^1-y^1}) \, .
\ee
Note that while writing the second line of Eq. (\ref{8.10}), we have used the 
Cauchy-Riemann equations
\be\label{7a}
\varepsilon^{ij}{\partial\over\partial x^j} \ln\mid {\bf x}-{\bf y} \mid 
\ = - {\partial\over \partial x^i} \phi ({\bf x}-{\bf y}) \, .
\ee

It is worth pointing out that
$\varepsilon^{ij}{\partial\over\partial x^j} G( {\bf x}-{\bf y} )$
is ill-defined at ${\bf x}={\bf y}$. Thus one has to supplement 
Eqs. (\ref{8.8}) and (\ref{8.9})
with a regularization prescription. One such prescription is 
\be\label{8b} 
\varepsilon^{ij}{\partial\over\partial x^j} G({\bf x})\longrightarrow
\varepsilon^{ij}{\partial\over\partial x^j} G^a({\bf x})
\ee
where the regulated Green function $G^{(a)}({\bf x})$ is
\be\label{8c} 
G^{(a)}({\bf x}) = {1\over a\pi} \int d^2 y ({1\over 2\pi} 
\ln \mid {\bf x- y} \mid) e^{-y^2/a} \, .
\ee
This has the desired property that
\be\label{8d}  
\lim_{a\rightarrow 0}  G^{(a)}({\bf x}) = G({\bf x}) 
= {1\over 2\pi} \ln \mid {\bf x} \mid
\ee
while for any {\it a}
\be\label{8e} 
\lim_{x\rightarrow 0} \varepsilon^{ij} {\partial\over\partial x^j} G^{(a)}({\bf x}) = 0
\ee
so that once Eq. (\ref{8c}) is systematically used, all ambiguities are 
eliminated.

If one is now allowed to move the derivative operator outside the integral 
(\ref{8.10}), then 
one could express ${\bf A}$ as a gradient. However, $\phi ({\bf x}-{\bf y})$ 
is a multi valued function. 
Hence one must first fix a branch-cut in the $y$-plane starting at $x$ so as to
make it single-valued. No matter what choice is made for this cut, the 
resulting
range of integration of ${\bf y}$ will depend on ${\bf x}$ and hence extra 
contributions
are produced in moving $\partial/\partial x^i$ outside the ${\bf y}$ integral. 
Thus, in general one can not write 
\be\label{8.12}
{\bf A}({\bf x}) = - {e\over 2\pi\mu}\nabla_x \bigg [\int d^2 y 
\phi ({\bf x}-{\bf y})\rho (y) \bigg ] \, .
\ee
so that in general ${\bf A}$ is not a pure gauge and hence it cannot be
removed by a gauge transformation. However, in the special case when 
$\rho (y)$ is a sum
of $\delta$-functions, ${\bf A}(x)$  
is indeed a pure gauge. Such a situation arises in the case of non-relativistic 
localized
point particles \cite{8le}. Let us assume that in the context of 
our non-relativistic model (\ref{8.1}) too, $\rho(y)$ is a sum of 
$\delta$-functions
in which case the CS gauge field $A_{\mu}$
is entirely determined by the matter configuration i.e. $\rho$ and ${\bf J}$.
 
Thus, in the case of localized densities, $A_{\mu}(x) 
= -\partial_{\mu}\Lambda (x)$ i.e.
the CS field is a pure gauge and hence it can be removed by the gauge 
transformation 
$A_{\mu}\longrightarrow A_{\mu}' = A_{\mu} +\partial_{\mu} \Lambda = 0$.
Thus, under such a singular transformation, covariant derivatives turn into 
ordinary derivatives, and the action (\ref{8.1}) becomes 
\be\label{8.20}
S' = \int d^3 x \bigg [ i\tilde\psi^+ \partial_0\tilde\psi+{1\over 2m}\tilde\psi^+
(\partial^2_1+\partial^2_2)\tilde\psi \bigg ]
\ee
where the new matter field $\tilde\psi$ is defined as
\be\label{8.21}
\tilde \psi (x) = e^{-ie\Lambda(x)}\psi(x) \, , \ \tilde\psi^+(x)
=\psi^+(x)e^{ie\Lambda(x)} \, .
\ee
The above action (\ref{8.20}) is that of a free, complex, non-relativistic, 
scalar
field $\tilde {\psi}$. However, we now show that such a field {\it does not} 
obey the conventional commutation relations as satisfied by $\psi$.

We can quantize the action (\ref{8.1})  
by imposing the equal-time commutation relations for the bosonic field $\psi$
\be\label{8.22}
[\psi({\bf x},t), \ \psi^+({\bf y},t)] =\delta({\bf x}-{\bf y})
\ee 
\be\label{8.23}
[\psi({\bf x},t), \ \psi({\bf y},t)] =0= [\psi^+({\bf x},t), 
\ \psi^+({\bf y},t)] \, .
\ee 
Since the gauge field ${\bf A}$ is a function of the number density operator 
$\rho (=\psi^+\psi),$
hence the commutator of ${\bf A}$ and $\psi$ is not trivial. In fact using 
Eqs. (\ref{8.10}) and (\ref{8.22}) we obtain
\be\label{8.24}
[ A^i({\bf x},t), \psi({\bf y},t) ] = - {e\over\mu} \varepsilon^{ij} 
{\partial\over\partial x^j} G({\bf x}-{\bf y})\psi ({\bf y}) \, .
\ee
On using the regularized Green function as given by Eq. (\ref{8c}), 
it then follows by using Eq. (\ref{8e}) that 
$[ A^i({\bf x},t), \psi({\bf x},t) ] =0$.
This is interesting because it means that there are no ordering ambiguities 
in the quantum theory as given by Eq. (\ref{8.1}).
 
One can now show that when $\psi$ obeys ordinary commutation 
relations,
$\tilde {\psi}$ obeys 
\be\label{8.29}
\tilde {\psi} ({\bf x},t)\tilde {\psi} (\bf y,t) = e^{i\pi\alpha}
\tilde {\psi} ({\bf y},t)\tilde {\psi} ({\bf x},t)
\ee
i.e. the matter field $\tilde {\psi}$ obeys anyonic commutation relations of 
statistics
$\alpha \ (= e^2/2\pi\mu)$. If instead, we make a cut along the negative 
$x'$-axis, then we
would obtain 
a phase factor $(e^{-i\pi\alpha})$, opposite to that in Eq. (\ref{8.29}).
Proceeding in the same way, it is easily shown that
if ${\bf x}\neq  {\bf y}$ then 
\be\label{8.31}
\tilde\psi ({\bf x},t)\tilde\psi^+({\bf y},t) = e^{-i\pi\alpha}
\tilde\psi^+({\bf y},t)\tilde\psi({\bf x},t) .
\ee
It must however be noted that for ${\bf x} = {\bf y}$, the 
phase proportional to 
$\alpha$ vanishes
and hence the canonical commutation relations remain unchanged. 

Some clarification is called for at this stage. What one has shown is that
the fields $\tilde\psi({\bf x},t), \tilde\psi({\bf y},t)$ satisfy anyonic
commutation relations with the phase factor $e^{+i\pi\alpha}$ or 
$e^{-i\pi\alpha}$ depending on how we make the cut. However, this is not
enough. What is really required is that the phase of the wave function 
changes both by $+\pi\alpha$ and $-\pi\alpha$ in response to which way we
braid in interchanging ${\bf x}$ and ${\bf y}$. No one has been able to show
this so far.
In fact, what we have shown above is the best that one can 
achieve
with local operators $\tilde {\psi},\tilde {\psi}^+$. Local information,  
like initial and final positions of particles, is simply not sufficient
to code the braiding, where we also have to specify which way the 
particles passed around each other in 
interchanging their positions. As I see it, the only way to take
care of this problem in this formalism is to choose such a definition of the
multi-valued 
function $\phi$ which will make $\tilde {\psi}$ a non-local operator.

Summarizing, it appears that within the non-relativistic field theory formalism,
anyons can only be described by non-local operators, which are hard to deal 
with.
If one insists on a local formulation, then one has to hide the statistics in 
an interaction with a CS field.
        
There is no doubt that ideally the  various effects of fractional spin, such as
the spin-statistics theorem should be understood only in a full fledged 
relativistic
quantum field theory. However, relatively little is known in this respect. 
The point is, if the fundamental fields are to carry 
fractional
spin, they must carry a multi-valued irreducible representation of $SO(2,1)$. 
This is because, a rotation of $2\pi$ does not leave the Wave function 
invariant, but rather, it multiplies it by a phase $e^{2i\pi j}$.  
We then
have the following two options.

The first option is that we define infinite component fields and 
from them construct 
one particle dynamics by imposing equations of motion that satisfy the 
requirement
that one-particle states provide multi-valued Poincar\'e 
equations. The most difficult part is the derivation of an action that 
reproduces these equations of motion. 
This requires handling a nonlocal theory and no one really
knows how to quantize such a theory.

The second option is to work with multi-valued fields by 
adding the
CS term to the action and essentially repeat what we have done above 
for the non-relativistic case.
Thus, instead of the non-relativistic model (\ref{8.1}), one could consider a 
relativistic
field theory, say a complex scalar field theory, coupled to an Abelian gauge 
field
with a CS kinetic energy term  (and no Maxwell term). 
Coming back to complex fields, 
one again 
wants to know if one can
construct local quantum field theory where the fundamental fields  
represent the creation and annihilation of anyons. On proceeding exactly as in 
the nonrelativistic case, 
one again obtains Eq. (\ref{8.6}).   
However, now the particles are not point
particles but are extended objects,
hence $\rho(y)$
cannot be a sum of delta functions. 
Thus it is not possible to 
write  ${\bf A}$ as a pure gauge and hence it cannot be 
removed
by a gauge transformation. 
Thus, it is not at all clear 
whether in the
relativistic case the only effect of the gauge field is to endow the particle
with arbitrary spin or if residual interactions are also 
present. A similar problem 
also
arises in models which emerge from the relativistic theory in the 
non-relativistic limit.
In particular, one obtains different results depending on which limit is taken 
first
i.e. the size of the extended object going to zero vis-a-vis the regulator 
parameter going to zero. Attempts have been made to tackle these 
problems by quantizing the theory with CS term on a lattice 
with or without the Maxwell term. So far, 
these attempts have met with only a limited success.

Thus it is fair to say that, so far we do not have a model in relativistic local
quantum field theory where the fundamental (non-interacting) field quanta are 
themselves anyons. In fact 
it appears unlikely that one can obtain a simple, 
local
(relativistic) Lagrangian for anyons. This is because, even in $2+1$ 
dimensions, spin
has to be an integer or half-integer for local fields. On the other hand,
fractional spin is 
admissible for fields which carry charges associated with
gauge symmetries (with accompanying flux integrals at infinity) which are 
typically
localizable only in 
space-like cones \cite{7bf,7fm}. This is what happens for 
example, when one
generates fractional 
spin by coupling point particles to a CS gauge 
field  
which has non-trivial long-ranged properties.

\end{document}